\documentclass{emulateapj}
\shorttitle{Non-barotropic RWI in 3D}
\shortauthors{M-K. Lin}

\usepackage{amsmath}
\usepackage{bm}
\newcommand{\p}{\partial}
\newcommand{\lmax}{l_\mathrm{max}}
\newcommand{\sbar}{\bar{\sigma}} 
\newcommand{\imgi}{\mathrm{i}} 
\newcommand{\ii}{\mathrm{i}}

\newcommand{\avg}[1]{\langle{#1}\rangle}

\newcommand{\dd}{\delta}
\newcommand{\real}{\operatorname{Re}}

\newcommand{\wbar}{\tilde{W}}
\newcommand{\qbar}{\tilde{Q}}
\newcommand{\dbigH}{\frac{H^\prime}{H}}
\newcommand{\Rp}{\frac{\rho_0^\prime}{\rho_0}}
\newcommand{\Fp}{\frac{g^\prime}{g}}
\newcommand{\Rpp}{\frac{\rho_0^{\prime\prime}}{\rho_0}}
\newcommand{\Fpp}{\frac{g^{\prime\prime}}{g}}
\newcommand{\ils}{L_s^{-1}}

\newcommand{\ihs}{H_s^{-1}}
\newcommand{\ciso}{c_\mathrm{iso}}
\newcommand{\hiso}{h_\mathrm{iso}}
\newcommand{\Hiso}{H_\mathrm{iso}}
\newcommand{\zeus}{\texttt{ZEUS-MP }}
\newcommand{\rsph}{r_\mathrm{sph}}
\defcitealias{lin12}{L12} 

\begin{document}

\title{Non-barotropic linear Rossby wave instability in 
three-dimensional disks}  

\author{Min-Kai Lin}
\affil{Canadian Institute for Theoretical Astrophysics,
60 St. George Street, Toronto, ON, M5S 3H8, Canada}
\email{mklin924@cita.utoronto.ca}

\begin{abstract}
Astrophysical disks with localized radial structure, such as protoplanetary disks
containing dead zones or gaps due to disk-planet interaction, may be
subject to the non-axisymmetric Rossby wave instability (RWI) that
lead to vortex-formation. The linear instability has recently been demonstrated
in three-dimensional (3D) barotropic disks. It is the purpose of 
this study to generalize the 3D linear problem to include an energy
equation, thereby accounting for baroclinity in three-dimensions.   
Linear stability calculations are presented for radially structured, 
vertically stratified, geometrically-thin disks with non-uniform
entropy distribution in both directions. Polytropic equilibria are
considered but adiabatic perturbations assumed. The unperturbed disk has a 
localized radial density bump making it susceptible to the RWI.  
The linearized fluid equations are solved numerically as a partial differential equation
eigenvalue problem. Emphasis on the ease of method
  implementation is given. It is found that when the polytropic index is fixed
and adiabatic index increased,   
non-uniform entropy has negligible effect on the RWI growth rate, but pressure and density 
perturbation magnitudes near a pressure enhancement increases away from the midplane. 
The associated meridional flow is also 
qualitatively changed from homentropic calculations. Meridional vortical motion
is identified in the nonhomentropic linear solution, as well as in a
nonlinear global hydrodynamic simulation of the RWI
in an initially isothermal disk evolved adiabatically. Numerical results suggest 
buoyancy forces play an important role in the internal flow of Rossby vortices. 
\end{abstract}

\section{Introduction}
%motivation
Understanding the stability and evolution of radially structured disks
is important for several astrophysical applications. Protoplanetary
disks are likely to have complex radial structure
\citep{terquem08,armitage11} such as the radial boundary
between magnetically active and inactive regions of the disk
\citep[`dead zones', ][]{gammie96}, and edges of gaps induced by a
giant planet \citep{lin86}. %These locations involve local 
%radial variations in the disk profile. 

Local variations in the disk profile, which both of the above examples
involve, are vulnerable to the so-called Rossby wave instability 
\citep[RWI,][]{lovelace99,li00,li01}. The RWI is a linear shear instability
associated with an extremum in the potential vorticity profile of the
disk, or a generalization thereof, and leads to local vortex 
formation in the nonlinear regime. This has been verified for both
dead zone boundary and planetary gaps in two-dimensional (2D) disks \citep[e.g.][]{varniere06,
  lyra08,lyra09, li09, lin11a, crespe11}.  

Previous studies have shown that disk vortices are able to concentrate dust particles, 
potentially assisting planetesimal formation \citep{barge95,inaba06}, which is of course
crucial for planet formation. They can also interact strongly with planets, leading to non-monotonic orbital 
migration \citep{yu10,lin10}. Although protoplanetary disks are thin, they are nevertheless 
three-dimensional (3D), so modeling these processes in 3D is necessary. 
 
Recently, the RWI has been demonstrated in 3D 
geometry in the context of protoplanetary disks
\citep{meheut10,meheut12b,meheut12c,meheut12,umurhan10, lin12, lin12b,lin12c,
  lyra12}. These models have, however, employed a barotropic or 
nearly-barotropic equation of state. They can therefore be regarded as
the thin-disk version of the Papaloizou-Pringle instability
\citep[PPI,][]{papaloizou84,papaloizou85,papaloizou87,goldreich86,narayan87},
originally  discovered for 3D pressure-supported thick  
tori. It is clearly of interest to extend 3D RWI calculations to
non-barotropic flow, which was one of the features that
distinguished the 2D RWI from the original PPI \citep{lovelace99}. 

Given that the RWI and PPI involve the same physics, that is, wave-coupling across 
co-rotation \citep{goldreich86,umurhan10},
it is worth pointing out that the PPI 
has in fact been generalized to nonhomentropic tori. \cite{frank88}
found that entropy gradients did not significantly affect instability
growth rates, %\footnote{They also considered \emph{baroclinic} tori.} 
while \cite{kojima89} concluded non-uniformity in entropy has
similar effects as compressibility. They also found perturbations have
weak vertical dependence, in agreement with analytical arguments for
%based on?
homentropic flow \citep{papaloizou85,goldreich86}.  

In this work, we study what is essentially the nonhomentropic PPI in
rotationally-supported thin 3D disks, a geometry relevant to 
protoplanetary disks. This is equivalent to an extension of the 2D RWI
studies of \cite{lovelace99} to 3D, and we will adopt such nomenclature. 

We consider the problem in the linear regime. Although the role
of the RWI in protoplanetary disks must be  determined through
nonlinear hydrodynamic simulations, linear calculations are
nevertheless a useful way to study the instability at low computational cost. 
It is also important to have such calculations at hand for comparison with 
nonlinear simulations. 
%Indeed, this is the primary purpose of the present calculations.

Linear disturbances in 3D disks are governed by complicated partial
differential equations \citep{kato01}. Even with a numerical approach, 
computing unstable modes is no simple task. One method is to evolve 
the linear equations as an initial value problem
\citep{papaloizou87,frank88} and measure growth rates from data. For
special disk equilibria, one can convert the problem to a set of
ordinary differential equations
\citep[][hereafter \citetalias{lin12}]{tanaka02,zhang06,meheut12,lin12}, but the derivation of which can
be tedious. Thus, our study is also motivated by the desire to  reduce
this complexity when a numerical method is sought out.  

We pursue a numerical solution to the two-dimensional eigenvalue problem. 
This approach has been taken by \cite{kojima86,kojima89b} using finite-difference 
and finite-element methods. Inspired by the aforementioned studies, we employ finite differences
in the radial direction and a pseudo-spectral method to treat the vertical direction \citep{lin12c}. 
We formulate the linear problem with numerical implementation in mind, 
so that much of the algebra can be taken care of by the 
numerical scheme, should one choose to do so.   

%problem has also been solved in full, using finite differences or
%finite element methods \citep{kojima86,kojima89b}.

This paper is organized as follows. In \S\ref{setup} we list the governing
equations and describe the polytropic disk equilibria under consideration. 
The linear problem is defined in \S\ref{linear} 
and the numerical method stated in \S\ref{numerics}. 
Linear simulations are presented in \S\ref{simulations} for disks with moderate
values of the polytropic index. Disks with an isothermal background are considered in \S\ref{isothermal},
where a nonlinear hydrodynamic simulation is also described. We summarize in \S\ref{summary} with
a discussion of important caveats and possible extensions to this study.  

%previous works (RWI)
%previous works (PPI)
%previous works (numerical methods)
%this work
\section{Disk model}\label{setup}
We consider a non-self-gravitating, inviscid fluid disk orbiting a
central star of mass $M_*$ and adopt cylindrical co-ordinates
$(r,\phi,z)$ centered on the star. The system is governed by the Euler
equations: 

\begin{align}
  &\frac{\p\rho}{\p t} + \nabla\cdot(\rho\bm{v}) = 0, \label{mass_eq}\\
  &\frac{\p\bm{v}}{\p t} + \bm{v}\cdot\nabla\bm{v} =
  -\frac{1}{\rho}\nabla p -\nabla\Phi_*, \label{momentum_eq}\\
  &\frac{\p}{\p t}\ln{s} + \bm{v}\cdot\nabla\ln{s} = 0, \label{energy_eq}
\end{align}
where $\rho$ is the mass density, $\bm{v}$ is the velocity field, $p$
is the pressure and we refer to $s\equiv p/\rho^\gamma$ as the
entropy, %where $\gamma$ is the ratio of specific heats and
where the ratio of specific heats $\gamma$ is assumed constant. 
In the momentum equation, $\Phi_*$ is the gravitational potential of the 
central star. Eq. \ref{energy_eq} describes adiabatic evolution.   

A direct consequence of Eq. \ref{mass_eq} and Eq. \ref{momentum_eq} 
is an equation for the vortensity 
$\bm{\zeta}\equiv\nabla\times\bm{v}/\rho$,  
\begin{align}\label{pv}
  \frac{D\bm{\zeta}}{Dt} = \bm{\zeta}\cdot\nabla\bm{v} +
  \frac{1}{\rho^3}\nabla\rho\times\nabla p,
\end{align}
where $D/Dt\equiv\p_t + \bm{v}\cdot\nabla$ 
is the Lagrangian derivative. The second term on the RHS is the baroclinic 
vorticity source. It is absent in barotropic flow for which $p=p(\rho)$. 
In this work, we consider barotropic equilibria but generally
non-barotropic disturbances, so the baroclinic term is effective in the perturbed state.  

\subsection{Polytropic equilibrium}
The unperturbed disk is steady, axisymmetric and polytropic. That is  
\begin{align}\label{poly_eq}
  p = K\rho^{1+1/n},
\end{align}
where $K$ is a constant and $n$ is the polytropic index.
We adopt the thin disk approximation \citepalias{lin12}, so 
the density field has the simple form $\rho=\rho_0(r)(1-z^2/H^2)^n$,
where $\rho_0(r)$ is the midplane density and $H(r)$ is the 
disk thickness. $\rho_0$ is specified indirectly by imposing a
surface density profile $\Sigma\propto r^{-\alpha}B(r)$
where $B(r)$  
is a Gaussian bump at $r=r_0$ with amplitude $A>1$ and width $\Delta
r$ \citep{li00}. The aspect-ratio at $r_0$ is parametrized as
$h\equiv H(r_0)/r_0$.  

The unperturbed velocity field is $(v_r,v_\phi,v_z)=(0,r\Omega,0)$ with
$\Omega=\Omega(r)$ for barotropic equilibria and is given via 
centrifugal balance with gravity and pressure. Note that for a thin,
non-self-gravitating disk the angular velocity is nearly Keplerian, 
$\Omega\simeq\Omega_k\equiv\sqrt{GM_*/r^3}$ where $G$ is the
gravitational constant. 

The above setup is the same as in \citetalias{lin12}, and equations defining the
equilibrium are listed therein. The limit $n\to\infty$ corresponds to
isothermal equilibria, and is treated as a special case in
\S\ref{isothermal}.  

Polytropic equilibria are adopted for simplicity and to allow  
direct comparison with \citetalias{lin12}, which considered
homentropic flow where $\Gamma\equiv 1+1/n =\gamma$. Then 
Eq. \ref{poly_eq} holds in the perturbed disk, replacing 
Eq. \ref{energy_eq}. Setting $\gamma\neq\Gamma$ gives a nonhomentropic
disk.  

Following \cite{lovelace99}, it is convenient to define the
following length-scales 
\begin{align}
  &L_p  = \left(\frac{1}{\gamma}\frac{\p\ln{p}}{\p r}\right)^{-1},\,
  &H_p  = \left(\frac{1}{\gamma}\frac{\p\ln{p}}{\p z}\right)^{-1},\\
  &L_s  = \left(\frac{1}{\gamma}\frac{\p\ln{s}}{\p r}\right)^{-1},\,
  &H_s  = \left(\frac{1}{\gamma}\frac{\p\ln{s}}{\p z}\right)^{-1}.
\end{align}
These are, respectively, the pressure and entropy length-scales in the
radial and vertical directions, which depend on both $r$ and $z$. Note
that for polytropic equilibria, the entropy and pressure
length-scales only differ by a constant multiplicative factor. %retain
                                %different notation?

\subsection{Stability criteria} 
We consider disk equilibria satisfying 
the Solberg-Hoiland criteria for stability against axisymmetric
perturbations:
\begin{align}
  \kappa^2 + N_r^2+N_z^2 >0,\quad \kappa^2N_z^2>0,
\end{align}
where $\kappa^2=r^{-3}d(r^4\Omega^2)/dr$ is the square of the epicycle frequency and 
\begin{align}
  N_r^2 = -\frac{c_s^2}{L_pL_s}, \, N_z^2 = -\frac{c_s^2}{H_pH_s} 
\end{align}
are the radial and vertical buoyancy frequencies, respectively,
and $c_s=(\gamma p/\rho)^{1/2}$ is the adiabatic sound speed \citep{tassoul00}.  
We also define $N^2 \equiv N_r^2 + N_z^2$.  
%toomre parameter 

Our disk models satisfy the Rayleigh criterion 
$\kappa^2>0$, which limits the surface density bump amplitude\footnote{This also      
  means that, by rescaling the density field, we can always  make the 
  Toomre stability parameter $Q_T\equiv \bar{c}_s\kappa/\pi
  G\Sigma\gg1$, where $\bar{c}_s$ is a typical sound-speed,
  to satisfy the assumption of a non-self-gravitating disk.}.  
Then 
we require $N_z^2>0$, or stability against vertical convection, so
the disk should be sub-adiabatically  
stratified with $\Gamma<\gamma$. For rotationally supported thin
disks, $|N_r^2|\ll\kappa^2$ so the first Solberg-Hoiland
condition is generally satisfied regardless of the equation of state \citep{li00}. 
Note that $N_z^2$ increases with $z$, so we expect the disk to be more stable at larger
heights.   

\subsection{Instability criterion}
In the original 2D RWI calculations, 
\cite{lovelace99} found that when there is an extremum in the generalized vortensity profile 
$\eta(r)$, where 
\begin{align}\label{gen_vort}
  \eta = \frac{\kappa^2}{2\Omega\Sigma}\times\left(\frac{\Pi}{\Sigma^{\gamma_2}}\right)^{-2/\gamma_2},
\end{align}
the disk may be unstable to non-axisymmetric perturbations localized
about the extremum. Here, $\Pi\equiv\int^{\infty}_{-\infty} p dz$ is the
vertically integrated pressure and $\gamma_2$ is the adiabatic
index in the two-dimensional energy equation
$D(\Pi\Sigma^{-\gamma_2})/Dt=0$.    

To use Eq. \ref{gen_vort} in characterizing 3D disks, we  
use results from \cite{goldreich86} to relate $\gamma_2$ and $\gamma$. 
\citeauthor{goldreich86} studied linear disturbances in homentropic
slender tori with a polytropic equation of state 
(Eq. \ref{poly_eq}). Assuming vertical hydrostatic equilibrium, they
showed that the vertically integrated system has an effective
polytropic index of  
$n_2=n+1/2$. If $\gamma_2=1+1/n_2$ then 
%\begin{align}\label{gamma2}
$  \gamma_2=(3\gamma-1)/(\gamma+1)$. 
%\end{align}
This  relation has been used by other authors
\citep[e.g.][]{li00,klahr04}. The assumptions made by
\citeauthor{goldreich86} do not strictly apply to our case
(nonhomentropic equilibria and non-zero vertical motions) but their
result will suffice for diagnostic purposes.     

The polytropic disk equilibria have $\Pi\propto\rho_0^\Gamma H$ and $\Sigma\propto \rho_0
H$, so the above definition gives 
\begin{align*}
  \eta\propto
  \frac{\kappa^2}{2\Omega}\Sigma^{\left(1-2\Gamma/\gamma_2\right)}H^{2\left(\Gamma-1\right)/\gamma_2}.
\end{align*} 
For the adopted parameter values, a surface density bump corresponds
to a local minimum in the generalized vortensity, so that
$d\eta/dr\simeq0$ at $r=r_0$. This is also close to a local
$\mathrm{min}(\kappa^2)$. These minima act to `trap' disturbances,
leading to instability \citep{li00}.

\section{Linear problem}\label{linear}
We consider Eulerian perturbations to the above equilibrium in the form
$\real[\dd\rho(r,z)\exp{\imgi(m\phi + \sigma t)}]$ and similarly for other
fluid variables. Here, $m$ is the azimuthal wavenumber taken to be a
positive integer and $\sigma=-\omega -\imgi\nu$ is a complex
frequency, where $-\omega$ is the real mode frequency and $\nu$ is the
growth rate. The co-rotation radius $r_c$ of a mode is
such that $m\Omega(r_c)-\omega=0$, and the RWI is characterized by
$r_c\simeq r_0$. For clarity, hereafter we omit writing out the
time and azimuthal dependence explicitly. 

The goal is to obtain a partial differential equation (PDE) for the
quantity $W\equiv\dd p/\rho$. An explicit form of this equation is given by
\cite{kojima89}, but our priority is
the ease of solution implementation. By writing individual equations in
standard form --- a sum of coefficients multiplying differential
operators --- we can formulate the linear problem  with convenient
variables, then transform to the desired ones by redefining said
coefficients. These transformations can be done in the numerical
code. 

We begin by writing down the linearized equations in
terms of the intermediate variables $\wbar = \rho W$  and
$\qbar\equiv c_s^2\dd\rho$. The momentum equations give   
\begin{align}\
  &\rho\dd v_r = -\frac{\imgi}{D}\left(\sbar\frac{\p\wbar}{\p r} +
  \frac{2m\Omega}{r}\wbar\right) + \frac{\imgi\sbar}{L_pD}\qbar,\label{vr}\\
  &\rho\dd v_\phi =
  \frac{1}{D}\left(\frac{\kappa^2}{2\Omega}\frac{\p\wbar}{\p r} + 
  \frac{m\sbar}{r}\wbar\right) - \frac{\kappa^2}{2\Omega L_pD}\qbar,\label{vphi}\\
  &\rho\dd v_z =
  \frac{\imgi}{\sbar}\left(\frac{\p\wbar}{\p z}-\frac{\qbar}{H_p} \right),\label{vz} 
\end{align}
where $\sbar=\sigma + m\Omega$ is the shifted frequency and
$D=\kappa^2 - \sbar^2$. The linearized continuity equation is 
\begin{align}
  \imgi\sbar\frac{\qbar}{c_s^2} + \frac{1}{r}\frac{\p}{\p r}\left(r\rho\dd
  v_r\right) + \frac{\imgi m}{r}\rho\dd v_\phi + \frac{\p}{\p
    z}\left(\rho\dd v_z\right)=0, 
\end{align}
and the linearized energy equation is
\begin{align}\label{lin_energy}
  \imgi\sbar\left(\qbar - \wbar \right) =
  c_s^2\left[\frac{1}{L_s}\left(\rho \dd v_r\right) +
    \frac{1}{H_s}\left(\rho\dd v_z\right)\right].
\end{align}
Inserting the momentum equations into the continuity and energy
equations yield a pair of PDEs:
\begin{align}
&\frac{\sbar}{r}\frac{\p}{\p r}\left(\frac{r}{D}\frac{\p\wbar}{\p
  r}\right) -\frac{1}{\sbar}\frac{\p^2\wbar}{\p
  z^2} + \left[\frac{2m}{r}\frac{\p}{\p
    r}\left(\frac{\Omega}{D}\right) - \frac{\sbar
    m^2}{r^2D}\right]\wbar\notag\\
  &-\frac{\sbar}{r}\frac{\p}{\p r}\left(\frac{r\qbar}{L_pD}\right) +
  \frac{1}{\sbar}\frac{\p}{\p z}\left(\frac{\qbar}{H_p} \right) +
  \left[\frac{2m\Omega}{rL_pD}-\frac{\sbar}{c_s^2}\right]\qbar = 0,\label{contin_bar}\\
  &\frac{\sbar}{L_sD}\frac{\p\wbar}{\p r} - \frac{1}{\sbar
    H_s}\frac{\p\wbar}{\p z} +
  \left[\frac{2m\Omega}{rL_sD}-\frac{\sbar}{c_s^2}\right]\wbar
  \notag\\
  &+ \left[\sbar\left(\frac{1}{c_s^2} - \frac{1}{L_sL_pD}\right) +
    \frac{1}{\sbar H_s H_p} \right]\qbar = 0.\label{energy_bar}
\end{align}%have not assume polytropic basic state so far (so valid
           %for isothermal disk
Eq. \ref{contin_bar}---\ref{energy_bar} are the governing equations
for linear disturbances. 
%The co-rotation radius $r_c$ of a mode is
%such that $m\Omega(r_c)-\omega=0$, and the RWI is characterized by
%$r_c\simeq r_0$. 

Next, we transform to the co-ordinates $(R,Z)=(r,z/H)$ so that the
background disk structure is separable. For example, the
density field becomes $\rho=\rho_0(R)g(Z)$. Then the unperturbed disk
occupies a  rectangular domain since $g(\pm1)=0$. The governing
equations become    
\begin{align}
  &a_1 \frac{\p^2 \wbar}{\p R^2} + b_1\frac{\p^2\wbar}{\p Z\p R} +
  c_1\frac{\p^2\wbar}{\p Z^2} + d_1\frac{\p \wbar}{\p R} + e_1\frac{\p \wbar}{\p
    Z} + f_1\wbar\label{contin_coeff_bar} \notag\\
  &+\bar{d}_1\frac{\p \qbar}{\p R} + \bar{e}_1\frac{\p \qbar}{\p Z} +
  \bar{f}_1\qbar=0, \\
  &d_2\frac{\p \wbar}{\p R} + e_2\frac{\p \wbar}{\p Z} + f_2 \wbar +
  \bar{f}_2\qbar = 0. \label{energy_coeff_bar}
\end{align}
Explicit expressions for the coefficients are listed in Appendix
\ref{expressions}. We write the above PDE pair for $(\wbar,\qbar)$ as 
\begin{align}
  &V_1\wbar + \bar{V}_1\qbar = 0,\label{op_eq1}\\
  &V_2\wbar + \bar{V}_2\qbar = 0.\label{op_eq2}
\end{align}
Since $\bar{V}_2$ is a multiplicative factor, we can eliminate $\qbar$
between Eq. \ref{op_eq1}---\ref{op_eq2} to obtain an equation for
$\wbar$: 
\begin{align}
  \left[V_1 -
    \bar{V}_1\left(\bar{V}_2^{-1}V_2\right)\right]\wbar\equiv V\wbar =  0. 
\end{align}
%The operator $V$ has the same form as $V_1$. 
The operator $V$ is obtained by updating the coefficients of $V_1$, so
they have the same form. 
%That is, a sum of
%coefficients mutiplying differential operators. 
%amounts to modifying the coefficients of $V_1$ in
%Eq. \ref{contin_coeff_bar}. 
Finally, we substitute
$\wbar = \rho W$ to obtain  
\begin{align}
  UW = 0. 
\end{align}
Construction of $V$, and hence $U$, requires the evaluation of 
$\bar{V_1}\left(\bar{V}_2^{-1}V_2\right)$ which involves radial and 
vertical derivatives of the coefficients in 
Eq. \ref{energy_coeff_bar}. In Appendix \ref{alternative} we outline 
an alternative numerical approach which circumvents the algebra. (This  
appendix also includes relevant formulae to redefine the PDE
coefficients for the transformation $V\to U$.) 

%% The The transformation $V\to U$ essentially redefines the coefficients in
%% Eq. \ref{contin_coeff_bar} and \ref{energy_coeff_bar} and the
%% relevant formulae are included in Appendix 

The key dependent variable is $W$, but we also interpret results using
$Q\equiv\qbar/\rho$. We refer to $W$ and $Q$ as pressure and density
perturbations, respectively. Then the entropy perturbation is
naturally defined as 
\begin{align}
  S \equiv W - Q. 
\end{align}

\subsection{Boundary conditions}
We consider disturbances radially confined about the 
density bump at $r=r_0$, so the inner and outer disk boundaries play
no significant role \citep{umurhan10}. 
Hence, for simplicity we set $\p_RW=0$ at radial boundaries. 
%tests with reduced boundaries...

Pressure and density perturbations are assumed to be symmetric about
the disk midplane. Henceforth we consider $z\geq 0$ without loss of generality.  
The default upper disk boundary condition is 
vanishing Lagrangian pressure perturbation at $Z=Z_s$:
\begin{align}\label{vbc}
  \Delta p 
  &\equiv \dd p + \bm{\xi}\cdot\nabla p=0,
\end{align}
where $\bm{\xi}$ is the Lagrangian displacement ($\nabla$ 
refers to cylindrical co-ordinates). We call this the free boundary condition. 
The surface function $Z_s$ is assumed constant 
for simplicity. If $Z_s$ is the 
zero-pressure surface, then Eq. \ref{vbc} can be satisfied 
automatically provided the perturbations are regular there. In 
practice, though, we take $Z_s<1$ to avoid the disk surface 
\citep[where entropy and its derivatives diverge, ][]{zhuravlev07}. 
%the boundary condition  
%is imposed explicitly by replacing the governing equation with 
%Eq. \ref{vbc} at $Z=Z_s$. 
Note that Eq. \ref{vbc}, together with 
Eq. \ref{lin_energy}, imply $\Gamma Q=\gamma W$ at the upper 
boundary. 

In some cases we adopt a solid upper boundary:
\begin{align}\label{vbc2}
  \dd v_z = Z_s\frac{dH}{dr}\dd v_r, 
\end{align}
meaning no flow perpendicular to the boundary ($\dd v_\perp=0$), and occasionally
we set $\dd v_z=0$. Upper disk boundary conditions are imposed explicitly by replacing the
governing equation with Eq. \ref{vbc} or \ref{vbc2} at $Z=Z_s$.  

%Eq. \ref{vbc} is expressed in terms of
%$W$, which has a similar form to $V_2$. This equation then replaces
%the governing equation at $Z=Z_s$. 

\subsection{Baroclinity}\label{baroclinic_effects}
Before proceeding to solve the linear equations, it is useful to have a 
qualitative picture of the solution to aid us in checking results. The
main difference from \citetalias{lin12} is baroclinity. 
Here, we discuss expected effects of the baroclinic source term in
Eq. \ref{pv}. 

As we will often examine meridional flow, consider the azimuthal
component of Eq. \ref{pv}, which can source vortical motion in the 
$(r,z)$ plane. When linearized, this baroclinic source term becomes 
\begin{align}
  %\hat{\bm{\phi}}\cdot
  \frac{1}{\rho^3}\left(\nabla\rho\times\nabla p \right)_\phi
  &\to
  %\xrightarrow[\text{linearize}]{}
  \frac{1}{\rho^2}\left( \frac{1}{L_p}\frac{\p}{\p z} -  
  \frac{1}{H_p}\frac{\p}{\p r}\right)\left(\qbar -
  \frac{\gamma}{\Gamma}\wbar\right) \notag\\
%  &=\frac{1}{\rho}\left(\frac{\Gamma}{\gamma H}\frac{\rho_0^\prime}{\rho_0}
%    \frac{\p}{\p Z} - \frac{1}{H_p}\frac{\p}{\p R}\right)\bar{S},\notag\\
  & =\frac{\Gamma}{\gamma\rho H}\left[\, 
    \smash{\underbrace{\frac{\rho_0^\prime}{\rho_0}
    \frac{\p\bar{S}}{\p Z}}_\dagger} \!\!\!\phantom{\frac{1}{1}} +\phantom{\frac{1}{1}}\!\!\!
    \smash{\underbrace{\frac{2nZ}{(1-Z^2)}\frac{\p\bar{S}}{\p
          R}}_{\ddagger}}\,\right].\label{baroclinic}  
\end{align}
where $\bar{S}\equiv Q - \gamma W/\Gamma$ and $^\prime$ denotes
differentiation with respect to the argument. We have utilized the
barotropic background in obtaining Eq. \ref{baroclinic}. In the
discussion below, perturbations are regarded as real quantities. 

The RWI is characterized by non-axisymmetric pressure/density enhancements 
radially localized about the density bump \citep{li01}. Assuming this
is qualitatively unchanged in a nonhomentropic disk, let us denote   
the midplane co-ordinate of the center of one such enhancement as 
$(r_0,\phi_0)$. We will precisely define $\phi_0$ later. For now,  
consider the $(r,z)$ plane at fixed $\phi=\phi_0$ and about $r=r_0$.

Eq. \ref{baroclinic} shows that non-uniformity in $\bar{S}$ can cause vortical
motion in the meridional plane. The distribution of $\bar{S}(R,Z)$ at the chosen azimuth can be
anticipated as follows. Note that
\begin{align*} 
  \bar{S} = \left(1-\frac{\gamma}{\Gamma}\right)W - S. 
\end{align*}
We first deduce the sign of $\bar{S}(r_0,0)$. For a pressure 
enhancement, $W(r_0,0)>0$ and $(1-\gamma/\Gamma)W(r_0,0)<0$ 
because $\gamma>\Gamma$. To determine the sign of the local entropy perturbation, $S(r_0,0)$,
we recall the background entropy $ s \propto \rho^{\Gamma-\gamma}$ so a 
density bump at $r_0$ corresponds to an entropy dip there. Now, the RWI has caused a 
pressure/density enhancement at $(r_0,0)$. This can be achieved by moving  
fluid in the vicinity of $(r_0,0)$, which 
has higher entropy, toward $(r_0,0)$. Then the midplane 
Eulerian entropy perturbation at $r_0$ is positive, i.e. $S(r_0,0)>0$. Therefore
$\bar{S}(r_0,0)<0$. 

Next, the free boundary condition implies $\bar{S}(R,Z_s)=0$. So 
$\bar{S}(r_0,Z)$ varies from a negative value at the midplane to zero at the upper disk boundary. Then it is
reasonable to assume $\bar{S}(r_0,Z)\leq0$. (A similar argument can be made for the solid upper boundary.)
The perturbation magnitude $|\bar{S}|$ should also decrease radially away from $r_0$, 
because the RWI presents radially localized disturbances. 

A simple distribution to satisfy the above properties is for $\bar{S}$ to have 
a local minimum at $(r_0,0)$ and is negative or zero in this region. It is most negative at
$(r_0,0)$ and becomes less negative away from it. Then $\p_Z\bar{S}> 0$,  
and $\p_R\bar{S}\geq0$ ($\p_R\bar{S}\leq0$) for $R>r_0$ ($R<r_0$).

Consider regions radially away from $(r_0,0)$. 
From the argument above, $\bar{S}$ should be roughly two-dimensional ($\p_Z\ll 1$) 
away from its minimum at $(r_0,0)$. 
Then the sign of the baroclinic source (Eq. \ref{baroclinic}) 
is dictated by that of ($\ddagger$). Even if
$\p_Z=O(1)$ in these regions, we expect  $\p_R\sim H^{-1}$ for a radially
localized disturbance. Then the magnitude of ($\dagger$) relative to
($\ddagger$) is of order  $|H\rho_0^\prime/Z\rho_0|(1-Z^2)$,
which is small for the adopted disk models  (for $Z\neq0$). So away from $r_0$ and the midplane, the 
radial variation of $\bar{S}$ is more important than its vertical variation. 
Of course, this argument does not apply where $\p_R\bar{S}=0$, 
which occurs at $Z=Z_s$ and is expected close to $r=r_0$.

Under the above assumptions we anticipate that away from the
midplane but not very close to the upper disk boundary, the sign of the baroclinic
source term is determined by the radial derivative of $\bar{S}$, 
which is positive (negative) exterior
(interior) to $r_0$. Close to or at $r_0$, provided $\bar{S}$ varies more rapidly in 
the vertical direction than radial,  
the sign of the baroclinic source is the same as that of $\rho_0^\prime$,   
which is \emph{typically} negative, but not always, due to a density bump. 

\section{Numerical procedure}\label{numerics}
The operator $U$ can be written in the same form as $V_1$. A matrix
representation of such an operator is described in \cite{lin12c} 
where details are given. We summarize here the main steps.  

The radial co-ordinate is discretized into $N_R$ uniformly spaced 
grid points. Let $W_i(Z)\equiv W(R_i,Z)$ denote the solution along the 
vertical line $R=R_i$. We set 
\begin{align}\label{cheby_exp}
  W_i(Z)=\sum_{k=1}^{N_Z} w_{ki}\psi_k(Z/Z_s), 
\end{align}
where the basis functions $\psi_k=T_{2(k-1)}$ are even Chebyshev polynomials
of the first kind. $N_Z$ is the number of basis functions and 
the highest polynomial order is $\lmax = 2(N_Z-1)$.  

Radial derivatives in $UW$ are replaced by 
central finite differences, and we evaluate 
vertical derivatives exactly at the $N_Z$ non-negative Lobatto 
grid points of $T_{\lmax}(Z/Z_s)$. This procedure performs the
conversion 
\begin{align}
  UW =0\to \bm{U}\bm{w}=\bm{0},
\end{align}
where $\bm{U}$ is a $(N_RN_Z)\times(N_RN_Z)$ block tridiagonal matrix
and $\bm{w}$ is a vector storing the $N_RN_Z$ pseudo-spectral
coefficients $w_{ki}$.  

The numerical problem is a set of linear homogeneous equations, 
$\bm{U}(\sigma)\bm{w}=\bm{0}$. Non-trivial solutions exist if
$\mathrm{det}\,\bm{U}=0$. This is achieved by varying $\sigma$ using
Newton-Raphson iteration. We only accept solutions where the reciprocal
of the condition number of $\bm{U}$ is zero at machine precision. The same
method of solution was employed  in \citetalias{lin12}.

\subsection{Results visualization}\label{visual}
The pressure perturbation $W$ is constructed from the 
pseudo-spectral coefficients $w_{ki}$. We then calculate $Q$ from
Eq. \ref{energy_coeff_bar} and velocity perturbations from
Eq. \ref{vr}---\ref{vz}. 

%% For convenience, we will 
%% refer to $W$ and $Q$ as the pressure and density perturbations,
%% respectively. It is then natural to define the entropy perturbation as 
%% \begin{align}
%%   S\equiv W - Q. 
%% \end{align}

We examine real perturbations about the \emph{vortex core}
$(r,\phi)=(r_0,\phi_0)$, where $m\phi_0=-\arg[{W(r_0,0)}]$.  
Setting $\phi=\phi_0$ is equivalent to redefining a physical
perturbation as 
\begin{align}
 X\to\real[X(r,z)W^*(r_0,0)],
\end{align}
where $X$ represents $W,\, Q,\, S$ or $\dd\bm{v}$, and $^*$ denotes
complex conjugate. All perturbations are regarded as real hereafter.  
In practice $(r_0,\phi_0)$ is close to a local 
maximum of pressure perturbation.
The magnitude of $X$, as redefined above, is arbitrary but its sign is
not. 

As an empirical measure of flow three-dimensionality, we compare 
vertical and horizontal motions near the bump radius using
$\avg{\theta_m}$, where 
\begin{align} 
  \theta_m^2 =\frac{\dd v_z^2}{\dd v_r^2 + \dd v_z^2}, 
\end{align}
and $\avg{\cdot}$ denotes averaging over $R\in[0.8,1.2]r_0$ and
$Z\in[0,Z_s]$ at $\phi=\phi_0$. 

%% This choice of azimuth is motivated by non-linear simulations
%% which show that the RWI leads to vortex formation about the bump
%% radius $r_0$, and the vortices are associated with pressure
%% enhancements \citep{li01}. 

\section{Linear simulations}\label{simulations}
We adopt units such that $G=M_*=1$. Our main calculations are
summarized in Table \ref{linsims}. For these runs the computational
domain is $R\in[0.4,1.6]r_0$, $Z\in[0,Z_s]=[0,0.9]$, and 
$\alpha=0.5$ for the power-law part of the surface density 
profile. The bump radius, amplitude and width are set to $r_0=1$,
$A=1.4$ and $\Delta r=0.05r_0$, respectively. 
We consider modes with $m=3$ unless otherwise stated. 
Slightly different setups are employed in \S\ref{isothermal} to
explore the isothermal limit.   

The new parameter for nonhomentropic disks, 
compared to homentropic flow in \citetalias{lin12}, is the adiabatic index
$\gamma$. We therefore focus on examining the effect of entropy
gradients due to $\gamma\ne\Gamma$. Cases 0---4 have fixed polytropic index
$n=1.5$, and therefore identical background density and velocity profiles,
but variable adiabatic index $\gamma \geq 5/3$. 
Cases 5---8 have fixed adiabatic index $\gamma=1.4$, but variable polytropic 
index $n\geq 2.5$. 

%It should be noted that 
%§with other parameters fixed, increasing $n$ decreases the bump in disk thickness. 

\begin{figure}[!t]
  \centering
  \includegraphics[scale=.425,clip=true,trim=0cm 1.84cm 0cm
    0cm]{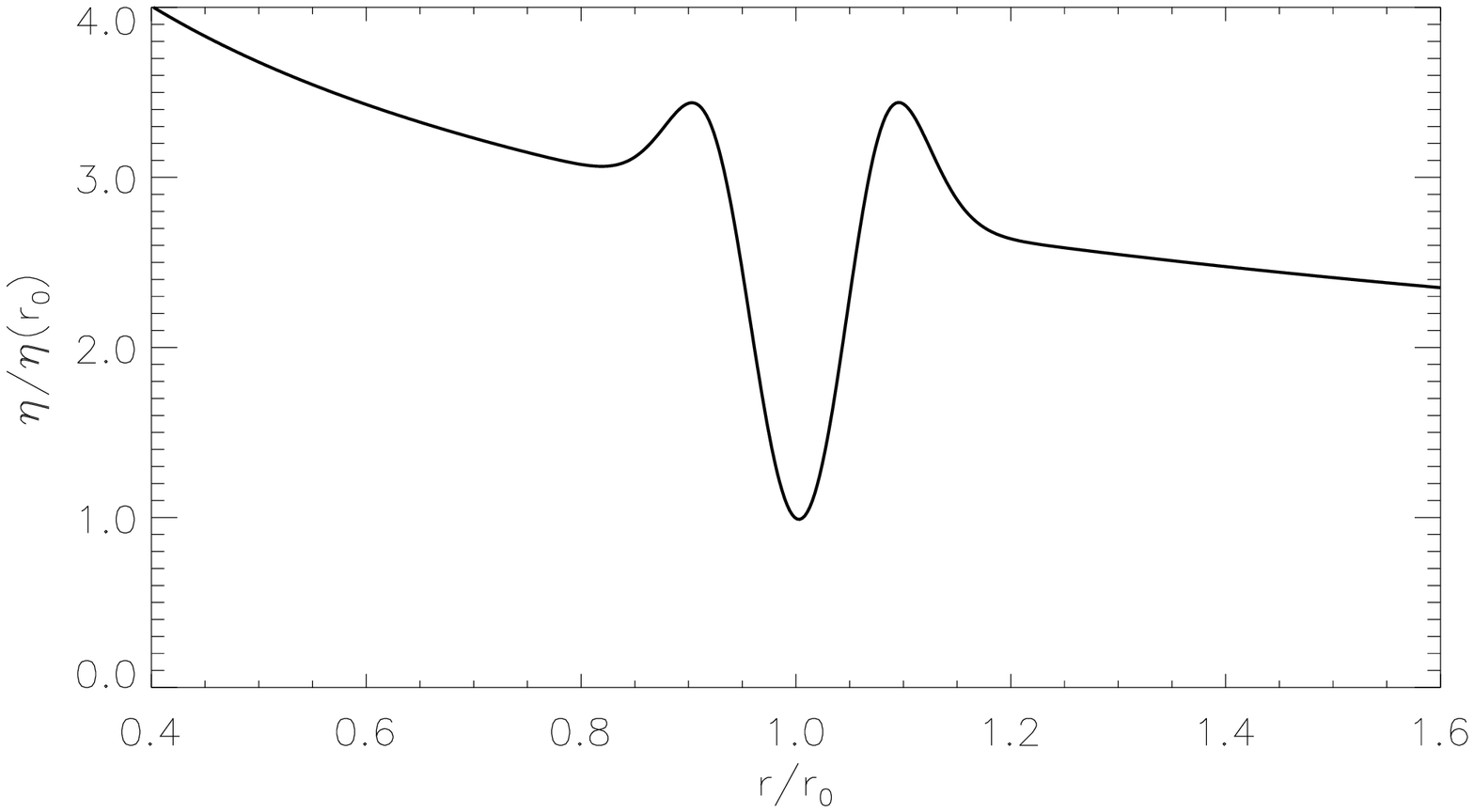}
  \includegraphics[scale=.425,clip=true,trim=0cm 0cm 0cm
    0.25cm]{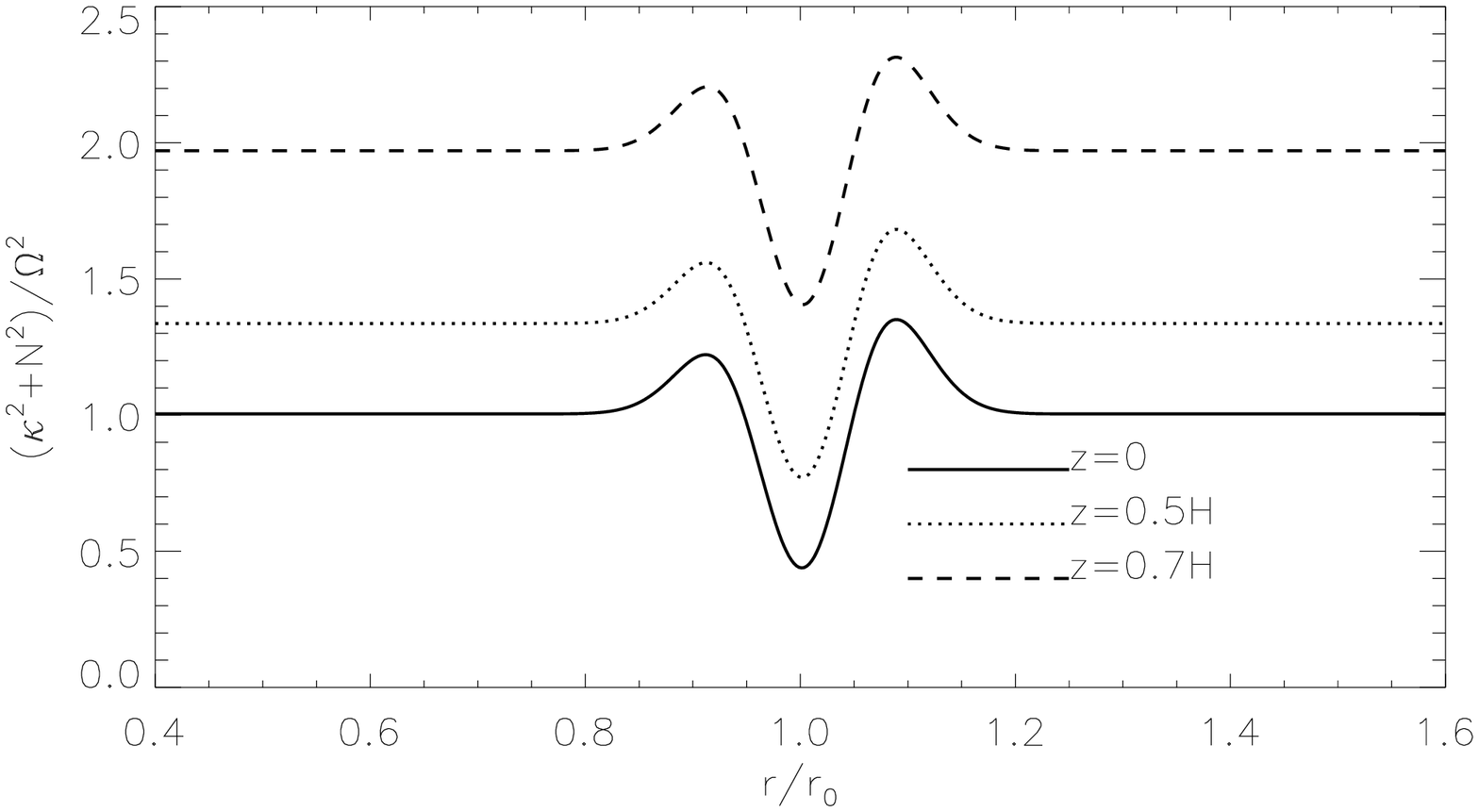}
  \caption{Equilibrium profile for a nonhomentropic disk with $n=1.5$ 
    and $\gamma=2.5$ (case 3). The generalized vortensity (top) 
    and $\kappa^2+N^2$ at three heights (bottom) are shown. 
    \label{fiducial_basic}}
\end{figure}

An example of nonhomentropic equilibrium, with $n=1.5$ ($\Gamma=5/3$)
and $\gamma=2.5$, is shown in Fig. \ref{fiducial_basic} (case 3).
The generalized vortensity and $\kappa^2+N^2$ are plotted. 
As expected for a density bump, the generalized vortensity has a local
minimum at $r=r_0$. It corresponds to $\mathrm{min}(\kappa^2/\Omega_k^2)=0.43$. 
The increase in $\kappa^2+N^2$ with respect to
height is due to $N_z^2$ (since $N_r\sim h N_z$ near the upper
boundary). Note that $N_z^2\gtrsim\Omega^2$ for
$|z|\gtrsim0.7H$ in this case.  

%generalized vortensity and ksq + nsq

%numerical parameters. we're not going to do detailed parameter study
%(done in L12)

The discretized problem is solved with standard matrix routines 
provided in the \texttt{LAPACK} package. The default resolution is
$(N_R,N_Z)=(512,12)$, corresponding to $\lmax=22$. 
%we have tested for convergence? wait for HR case

\begin{deluxetable}{llllccc}
  \tablecaption{Summary of main linear simulations\label{linsims}}
  \tablehead{\colhead{Case} &\colhead{$\Gamma$} & \colhead{$\gamma$}
    &\colhead{BC\tablenotemark{$\dagger$}}&\colhead{$\omega/m\Omega_0$} & \colhead{$10^2\nu/\Omega_0$} &
    \colhead{$\avg{\theta_m}$}  } 
  \startdata
  \multicolumn{7}{c}{$h=0.14$} \\
  \hline
  0 & 1.67 & 1.67 &$\Delta p=0$  &0.9941 & 10.74 & 0.33 \\ %theta at core [0.98,1.02] is 0.53
  1 & 1.67 & 1.8  &$\Delta p=0$  &0.9937 & 10.80 & 0.36 \\ %                             0.65
  2 & 1.67 & 2.0  &$\Delta p=0$  &0.9931 & 10.86 & 0.39 \\ %                             0.65 
  3a & 1.67 & 2.5 &$\Delta p=0$  &0.9919 & 10.99 & 0.44 \\ %                             0.58
  3b & 1.67 & 2.5 &$\dd v_\perp=0$  &0.9911 & 11.34 & 0.41 \\%                           0.52
  4 & 1.67 & 3.0  &$\Delta p=0$  & 0.9910      &  11.07     &  0.47    \\%               0.55
  \hline
  \multicolumn{7}{c}{$h=0.2$} \\
  \hline
  5 & 1.4 & 1.4  &$\Delta p=0$  & 0.9923 & 16.66 & 0.24 \\ %n=2.5, theta at core is 0.46
  6 & 1.33& 1.4  &$\Delta p=0$  & 0.9917 & 13.81 & 0.31  \\ %n=3.0,                 0.63
  7 & 1.29& 1.4  &$\Delta p=0$  & 0.9912 & 11.38 & 0.34  \\ %n=3.5,                 0.61  
  8 & 1.25& 1.4  &$\Delta p=0$  & 0.9909 & 9.246 & 0.36    %n=4.0,                 0.56 
  \enddata
   \tablenotetext{$\dagger$}{Boundary condition at $Z=Z_s$.}

%theta averaged between 0.9, 1.1 is
%case 0: 0.20
%case 1: 0.24
%case 2: 0.29
%case3a: 0.34
%case3b: 0.34
%case4 : 0.37
%
%
%case5: 0.18 
%case6: 0.26
%case7: 0.31
%case8: 0.34
\end{deluxetable}

%code test/gamma=Gamma case
\subsection{Homentropic reference case}
\begin{figure}[!t]
  \centering
  \includegraphics[scale=.425,clip=true,trim=0cm 1.cm 0cm
    0cm]{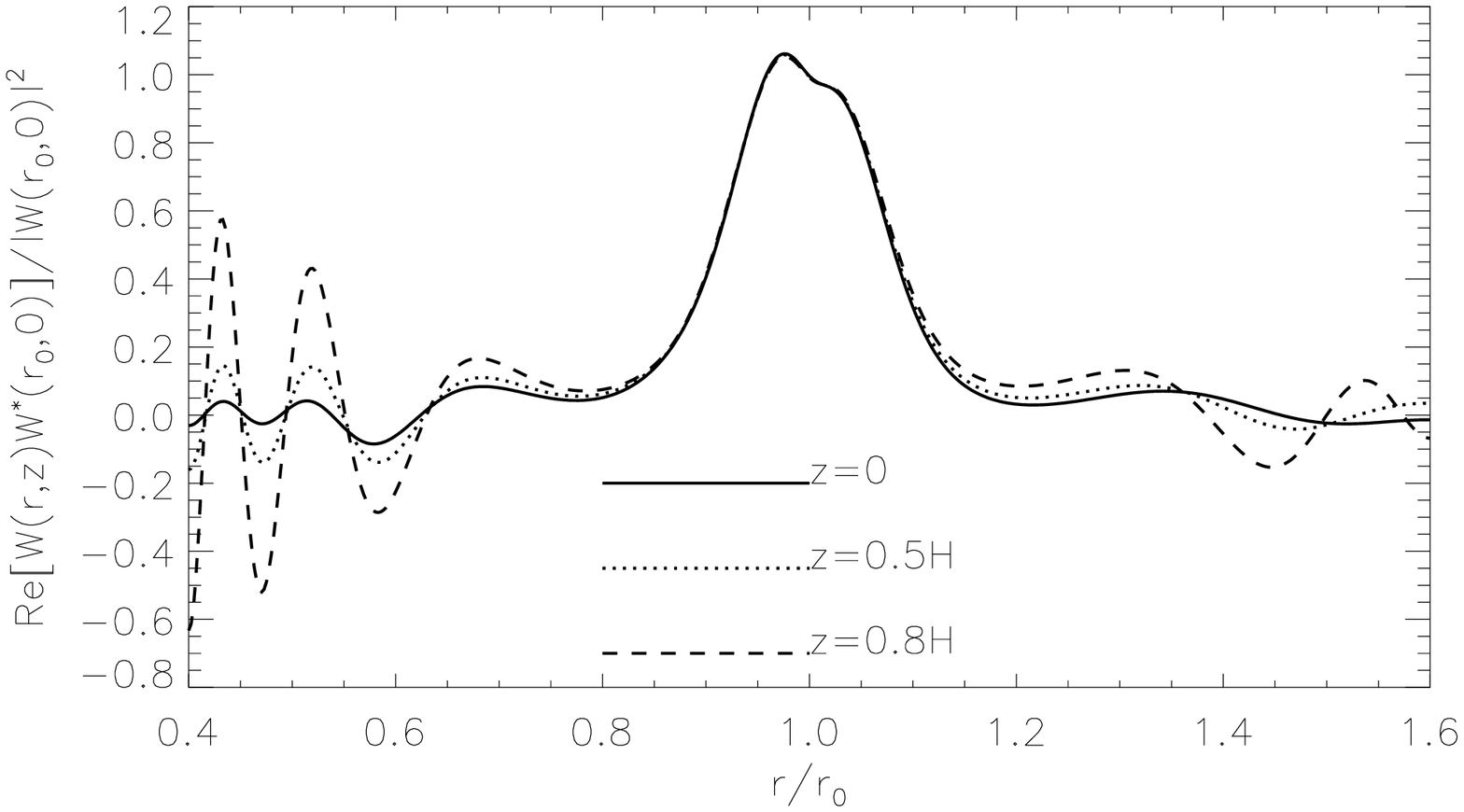}
  \includegraphics[scale=.425,clip=true,trim=0cm 0cm 0cm
    0.25cm]{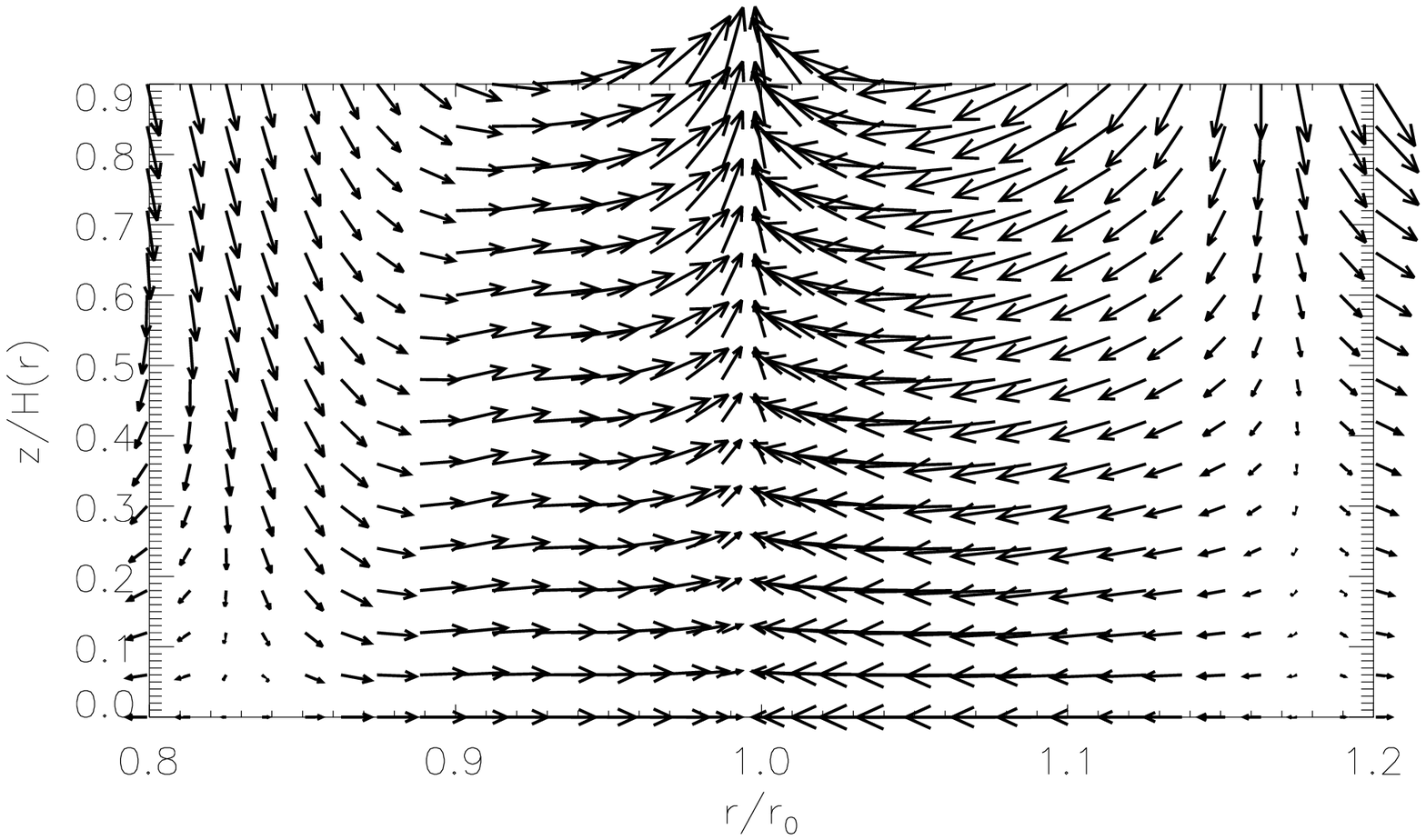}
  \caption{Real perturbations for the homentropic case 0
    ($\Gamma=\gamma=1.67$).  %in 
%    \citetalias{lin12} ($\Gamma=\gamma=5/3$). 
    The pressure perturbation $W$ at three heights (top) and
    meridional velocity perturbation (bottom) near the vortex core are
    shown.  
    \label{homentropic_case}}
\end{figure}

For comparison purposes, we reproduce the fiducial homentropic
calculation in \citetalias{lin12} by setting $\gamma=5/3$ (case
0). Then $W=Q$ since $\ils=\ihs\equiv0$ (Eq. \ref{lin_energy}). This also
serves as a test for our numerical method. The eigenfrequency and
perturbations shown in Table \ref{linsims} and
Fig. \ref{homentropic_case} agrees well with \citetalias{lin12}. In
co-rotation region $R\in[0.8,1.2]r_0$, $W$ is nearly independent
of height and the vortex core has upwards motion.     

%a non-homentropic case in detail
\subsection{Nonhomentropic example}
We now examine case 3a with $\gamma=2.5,\,\Gamma=1.67$. The
eigenfrequecy $\sigma$ is close to case 0, but the growth rate
is slightly larger in the nonhomentropic disk.    

\begin{figure}[!t]
  \centering
  \includegraphics[scale=.425,clip=true,trim=0cm 0.cm 0cm
    0cm]{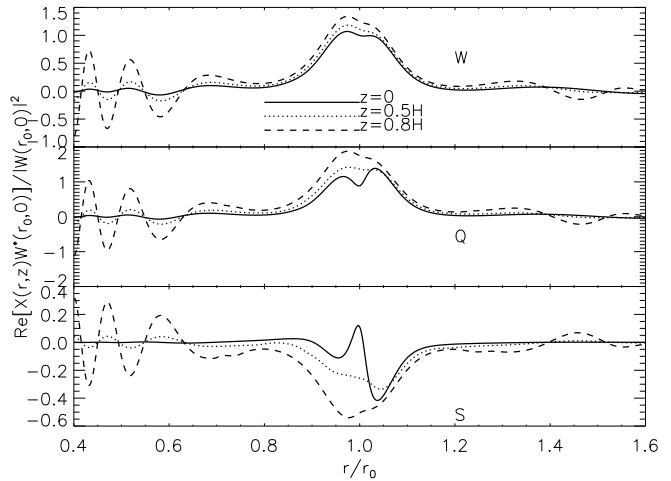}
  \caption{Pressure (top, $W$), density (middle, $Q$) and entropy
    (bottom, $S$) perturbations for the nonhomentropic case 3a 
    ($\Gamma=1.67,\,\gamma=2.5$).  
    \label{nonhomentropic_case}}
\end{figure}

Fig. \ref{nonhomentropic_case} shows the pressure, density and
entropy perturbations at several heights. Near $r_0$,  pressure
and density perturbations increase with height, unlike the homentropic case
where $W$ has weak $z$-dependence. Nonhomentropic disks  
generally have $W\ne Q$, as shown in
Fig. \ref{nonhomentropic_case}. The difference between $W$ and $Q$
at the midplane is due to background radial entropy gradients $\ils$ 
since $\ihs(r,0)=0$.  

At co-rotation,  the density perturbation $Q$ increases with height faster than the pressure
perturbation $W$, which  results in a negative entropy perturbation. This
is consistent with the requirement $S=(1-\gamma/\Gamma)W$ at the upper
disk boundary. It is clear that $S$ has a stronger vertical dependence
than either $W$ or $Q$. 

We might have expected the above result on physical grounds. The homentropic case indicate
upward motion at the vortex core. If a positive (stable) vertical 
entropy gradient is introduced, then a fluid element displaced upwards should increase its density compared to the 
surrounding background, i.e. $Q>0$, and this should become more positive with height because 
vertical velocities increase in magnitude with height. The pressure perturbation is not expected 
to change as rapidly, because the fluid element can establish pressure 
equilibrium with its surroundings. 

%, but $|S|$ is somewhat small compared to $W,\,Q$  at the vortex core.   
%1D plots

\subsubsection{Entropy perturbation} 
\begin{figure}[!t]
  \centering
  \includegraphics[scale=.425,clip=true,trim=0cm 0.0cm 0cm
    0cm]{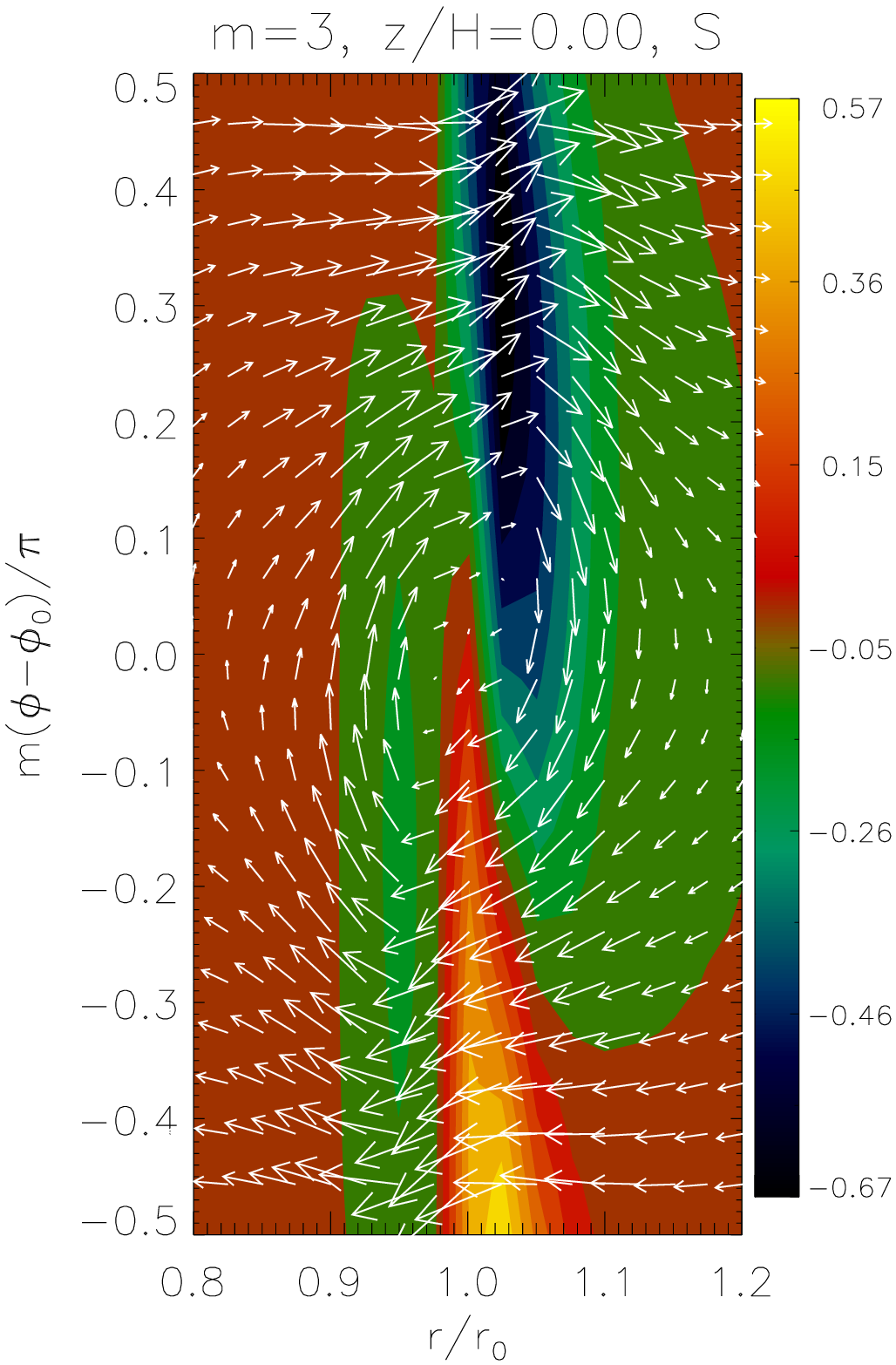}\includegraphics[scale=.425,clip=true,trim=2.2cm 
    0.0cm 0cm 0.0cm]{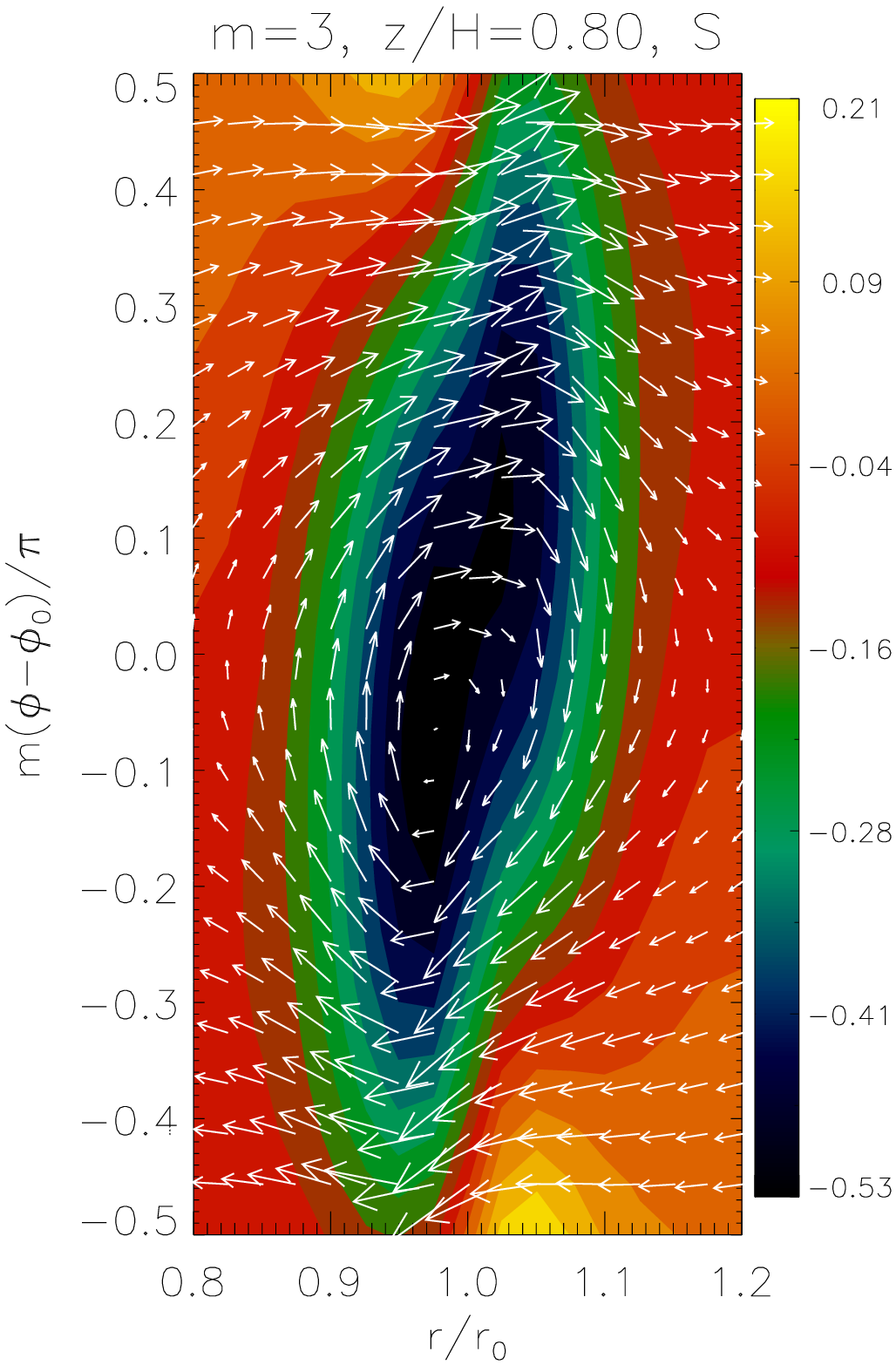}
  \caption{Entropy perturbation in midplane (left) and near the upper
    disk boundary (right) for the nonhomentropic case
    3a. Arrows show the perturbed velocity field projected onto this plane. 
  \label{2dflow}}
\end{figure}

We plot the entropy perturbation $S$ at $z=0$ and $z=0.8H$ in   
Fig. \ref{2dflow}. The figures are overlaid by the perturbed
horizontal flow, which are similar at both heights. The anti-cyclonic
flow pattern is commonly found in previous studies
\citep[e.g.][]{li00,li01}. Entropy gradients of this magnitude do not
affect this characteristic feature of the RWI.    

Therefore, we could have inferred some of the features in
Fig. \ref{2dflow} without solving the fluid equations {, by invoking entropy advection.} 
Consider the linearized energy equation near co-rotation where $\sbar \simeq
-\imgi\nu$ (which is close to $r_0$), 
%We can infer some of the features in Fig. \ref{2dflow} 
%% Near the co-rotation radius where $\sbar=-\imgi\nu$ (which is close
%% to the bump radius), the energy equation is approximately  
\begin{align}\label{entropy_gen}
  \dd s \sim -\nu^{-1}\dd\bm{v}\cdot\nabla s, 
\end{align}
and $\nu>0$ for a growing mode. Eq. \ref{entropy_gen} is only valid
within a small distance $\epsilon \ll \nu/|m\Omega_0^\prime|$ from
$r_0$. In this example, $\nu/|m\Omega_0^\prime|\simeq 0.02r_0$.  

%% The midplane entropy increases globally in the radial direction with a
%% dip at $r_0$,
The midplane entropy has a dip at the bump radius but it increases globally in the
radial direction, so $\p_rs>0$ at $r_0$.     
This naturally implies that inward (outward) radial flow  for
$\phi<\phi_0$ ($\phi>\phi_0$), i.e. anti-cyclonic motion, brings about
a local entropy increase (decrease) near $r_0$ at $z=0$. 
We recognize the qualitative similarity between the midplane flow pattern
in Fig. \ref{2dflow} and horseshoe turns induced by an embedded
planet. Entropy advection then leads to large radial entropy gradients
\citep{paardekooper10}, which can be seen in
Fig. \ref{nonhomentropic_case} on either side of $r_0$ at the
midplane. This gradient is, of course, growing exponentially in time, so it 
may be important even within the linear regime.  
%the linear baroclinic source for omega_z does not involve 
%radial derivative of ds (because it is multiplied by azi deriv of eqm pressure which is zero)
%but radial derivative of S will be involved in the nonlinear terms i.e.
% (phi deriv of dp) times (r deriv of s), and r deriv of s is large
%so even though we are working in the linear regime, this higher order nonlinear term (ignored in linear theory)  may 
%actually be significant because of the large r-deriv of s 

%the linear baroclinic source for omega_z involve
% (r deriv of p) times (azi deriv of ds)    and
% (r deriv of p) times (azi deriv of Q)     
%first term  might be large (Q ~ maximized at vortex core)
%ds changes sign across phi_0. so azi deriv of ds significant -> POSITIVE omega_z source (because r deriv of p usually negative). 

%That is, material of higher (lower) entropy moves to a region
%of lower (higher) background entropy.
%% Therefore, we could have used Eq. \ref{entropy_gen} to infer the local
%% entropy perturbation without solving the linear problem. Consider
%% first the midplane where only radial motions contribute to $\dd
%% s$. 
%% The perturbation is localized about an entropy dip at
%Away from the midplane we can assume vertical 
In the vertical dimension, if we assume the flow at
$(r_0,\phi_0)$ is unchanged from the homentropic case (i.e. upward),
then since the background entropy increases with height, the local
Eulerian entropy perturbation at the vortex core must become negative
away from the midplane, as observed.       

%vertical direction

\begin{figure}[!ht]
  \centering
  \includegraphics[scale=.425,clip=true,trim=0cm 0.cm 0cm
    0cm]{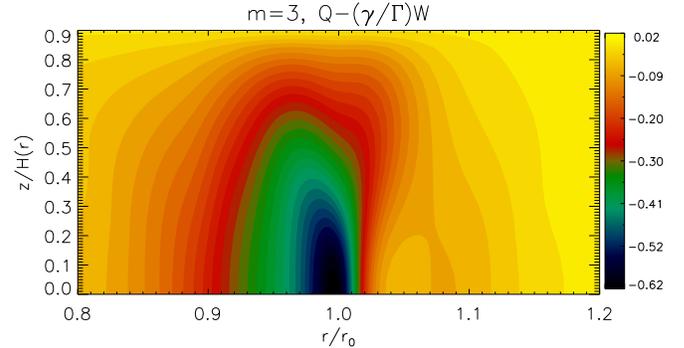}
  \caption{Map of the quantity $\bar{S}\equiv Q-\gamma W/\Gamma$ for
    case 3a (the real perturbation at $\phi=\phi_0$ is shown). $\bar{S}$
    appears in the baroclinic term in
    Eq. \ref{baroclinic}, as well as the expression for vertical velocity
    in Eq. \ref{core_dvz}.  The local minimum near $(r_0,0)$
    can be expected without solving the linear problem (see
    \S\ref{baroclinic_effects}). 
    \label{Sbar}}
\end{figure}

Related to the entropy perturbation is the quantity $\bar{S}\equiv
Q-\gamma W/\Gamma$. Its 
distribution shown in Fig. \ref{Sbar} agrees with expectations made in
\S\ref{baroclinic_effects}, namely it is mostly negative, with a local
minimum at the vortex core. 

\subsubsection{ Vertical vorticity perturbation }

\begin{figure}[!t]
  \centering
  \includegraphics[scale=.425,clip=true,trim=0cm 0.0cm 0cm
    0cm]{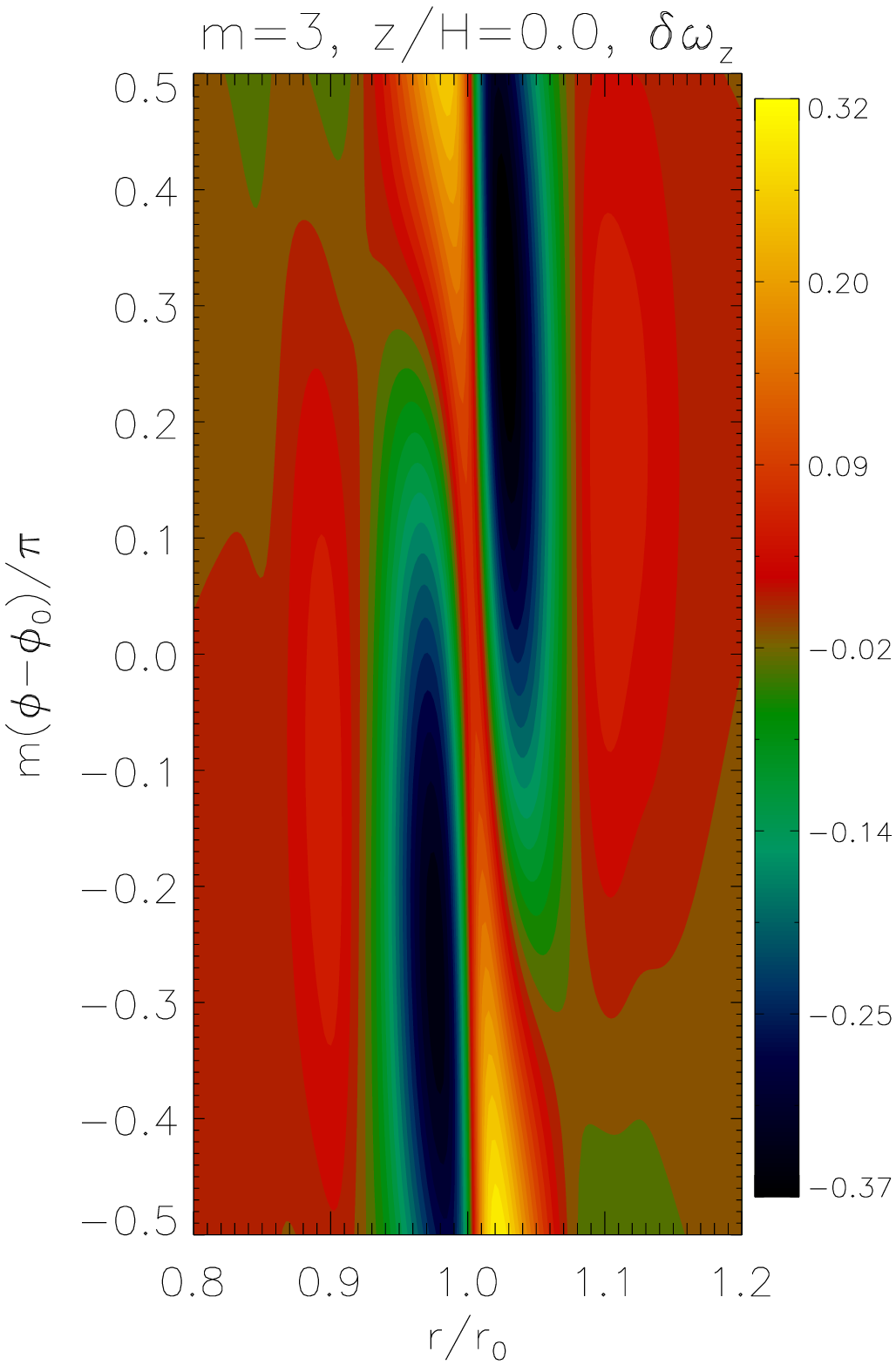}\includegraphics[scale=.425,clip=true,trim=2.2cm 
    0.0cm 0cm 0.0cm]{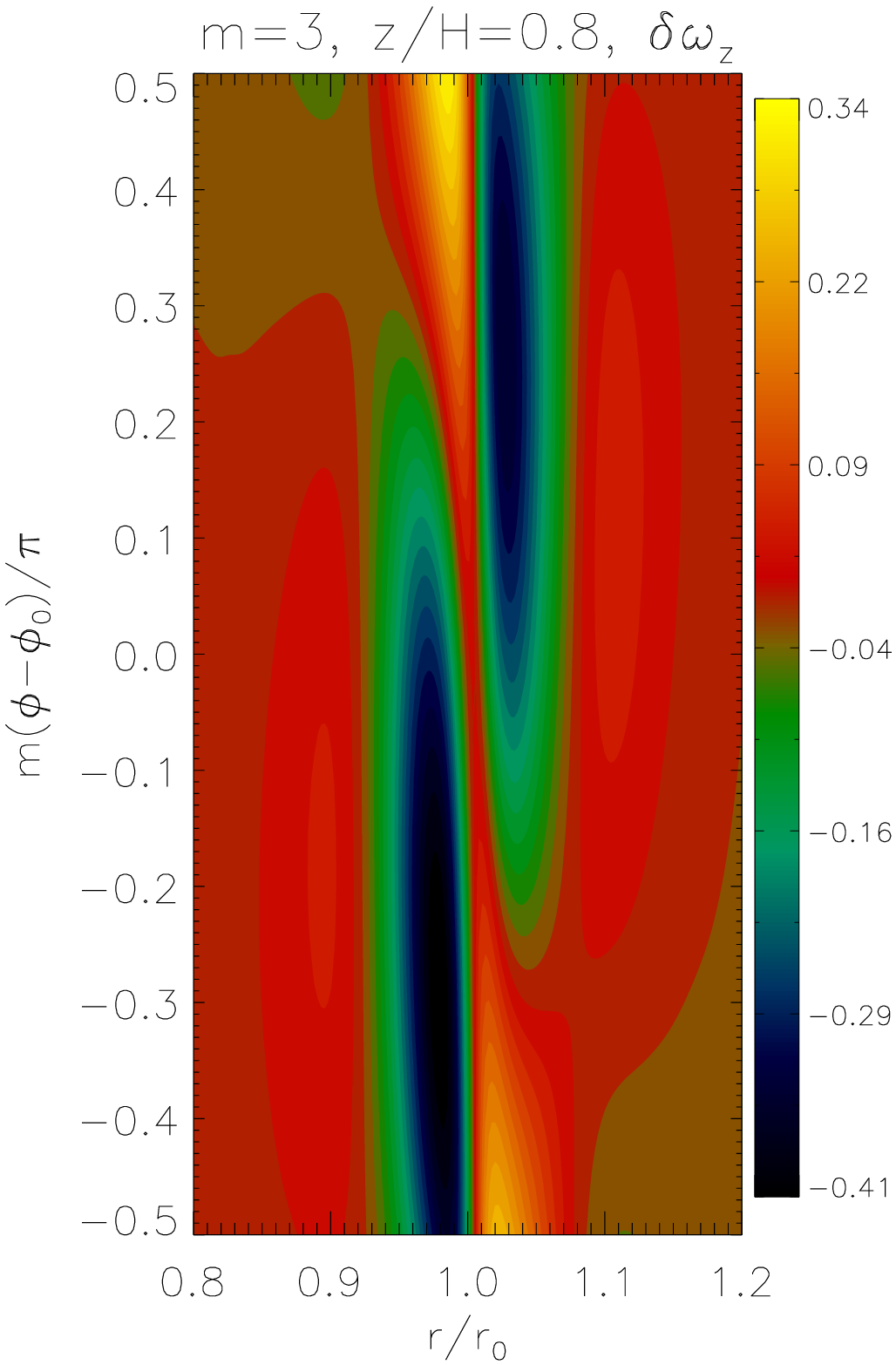}
  \caption{
      Perturbation to the vertical component of vorticity in the
      nonhomentropic case 3a at two heights in the horizontal
      plane. The vertical dependence is weak, but there is a slight
      increase in the maximum
      perturbation amplitude away from the midplane. 
      This figure is qualitatively similar to the top panel of Fig. 3 in
      \cite{meheut12b}. 
    \label{nonhomentropic_case_vortz_rp}}
\end{figure}

\begin{figure}[!t]
  \centering
  \includegraphics[scale=.425,clip=true,trim=0cm 0.cm 0cm
    0cm]{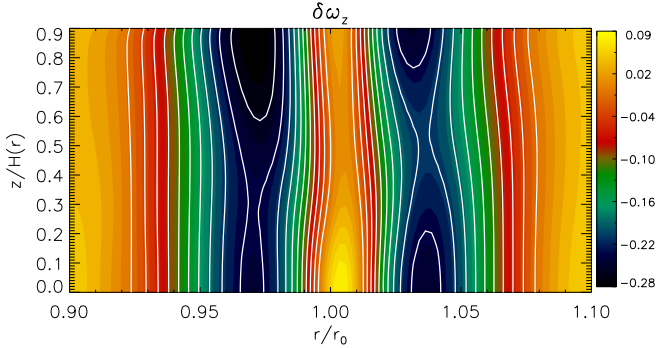}
  \caption{
      Perturbation to the vertical component of vorticity in the
      nonhomentropic case 3a, in the meridional plane at $\phi=\phi_0$. 
      Regions of $\delta\omega_z\leq0$ are
      delineated by white lines.
    \label{nonhomentropic_case_vortz}}
\end{figure}

Fig. \ref{nonhomentropic_case_vortz_rp}---\ref{nonhomentropic_case_vortz}
shows the perturbation 
to vertical vorticity,
$\delta\omega_z\equiv\hat{\bm{z}}\cdot\nabla\times\delta\bm{v}$, in
the horizontal and meridional planes, respectively. These plots 
agree with the identification of the linear RWI with a pair of edge-waves
propagating in the $\pm\phi$ directions along radial potential
vorticity gradients on either side of the bump radius $r_0$
\citep{umurhan10}. 

The background vorticity $\omega_z$ has a dip at
$r_0$. Then the positive/negative regions of $\delta\omega_z$ in
Fig. \ref{nonhomentropic_case_vortz_rp} is broadly consistent with the
advection of $\omega_z$ by the perturbed horizontal flow, in a similar
manner as the advection of entropy described in the previous section.   

Although the perturbed flow in the nonhomentropic case 
consists of vorticity columns (Fig. \ref{nonhomentropic_case_vortz}), 
there is actually a slight increase in $\mathrm{max}(|\delta\omega_z|)$ away from the midplane.     
This contrasts to \citeauthor{umurhan10}'s analytical model of the RWI in polytropic disks, 
where horizontal velocities, and hence $\delta\omega_z$, have no vertical dependence.

\subsubsection{ Meridional vortical flow and tilted vorticity columns}
\begin{figure}[!t]
  \centering
%  \includegraphics[scale=.425,clip=true,trim=0cm 0.cm 0cm
%    0cm]{n1.5_gmma2.5_vel3d_rz_dvz_nofill.ps}
  \includegraphics[scale=.425,clip=true,trim=0cm 0.cm 0cm
    0cm]{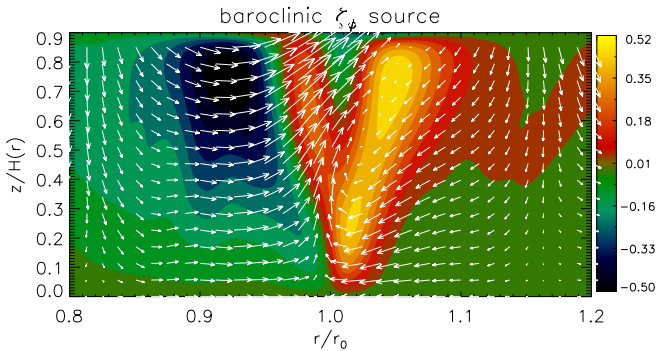}
  \caption{Meridional flow in the nonhomentropic case 3a. The azimuth
    taken for this slice is $\phi=\phi_0$. The contours
    show the baroclinic source term for azimuthal vortensity 
    (Eq. \ref{baroclinic} multiplied by $\rho$). The arrows show
    the perturbed velocity field projected onto this plane. 
    \label{nonhomentropic_case_dvz}}
\end{figure}

Fig. \ref{nonhomentropic_case_dvz} shows the perturbed velocity field
in the $(r,z)$ plane, with a map of the baroclinic source term defined
in \S\ref{baroclinic_effects}. The flow pattern is similar to the
homentropic case in that it is still converging toward $r_0$, and
vertical motion is predominantly upwards there. However, there is a
notable difference from the homentropic case --- vortical 
motion (of positive azimuthal vorticity) centered about 
$(r,z)=(1.02r_0,0.5H)$. It coincides with a region where the
azimuthal baroclinic source term is positive. Note that the sign of
the baroclinic source away from the midplane --- being positive
  (negative) for $r>r_0$ ($r<r_0$) --- is roughly consistent with
expectations made in 
\S\ref{baroclinic_effects}. 

%vortical motion sourced by baroclinity 

%velocity in rz plane

%% \subsubsubsection{Tilted voricity}

\begin{figure}[!t]
  \centering
  \includegraphics[scale=.425,clip=true,trim=0cm 0.cm 0cm
    0cm]{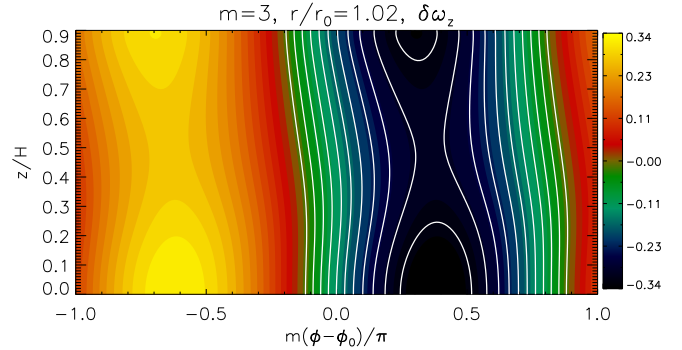}
  \caption{
    Perturbation to vertical vorticity in the $(\phi,z)$ plane
    at $r=1.02r_0$. Regions of $\delta\omega_z\leq0$ are delineated by
    white lines. The center of the meridional vortical motion
    identified in Fig. \ref{nonhomentropic_case_dvz} occurs at
    $(\phi,z)=(\phi_0,0.5H)$.  The azimuthal range $\phi-\phi_0\in[-0.5,0.5]\pi/m$
     corresponds to anti-cyclonic motion about the vortex core.
    \label{nonhomentropic_case_vort3d_pz}}
\end{figure}

  Vortical motion in the meridional plane also correlates to 
  misalignment between a column of negative vertical vorticity perturbation and  
  the vertical direction. This is demonstrated in 
    Fig. \ref{nonhomentropic_case_vort3d_pz} where 
    contours of $\delta\omega_z$ are shown to be 
    tilted in the $(\phi,z)$ plane. 
    We can quantify this tilt by
    calculating $1-\avg{\cos{\theta}}_Z$, where
    \begin{align*} 
      \cos{\theta} \equiv
      \left.|\p_\phi\delta\omega_z|\right/\left[(\p_Z\delta\omega_z)^2 + (\p_\phi\delta\omega_z)^2\right]^{1/2}
    \end{align*}
    and $\avg{\cdot}_Z$ denotes averaging over the vertical direction 
    at fixed $\phi=\phi_0$ shown in
    Fig. \ref{nonhomentropic_case_vort3d_pz}. 

    For the nonhomentropic case 3a, we find
    $1-~\avg{\cos{\theta}}_Z=0.011$. This value should be compared with the 
    homentropic case 0 where $1-~\avg{\cos{\theta}}_Z=~3.4\times10^{-5}$
    and the tilt is hardly noticeable.   
 
    We rationalize the small tilt observed in Fig. \ref{nonhomentropic_case_vort3d_pz} 
    by interpreting the nonhomentropic solution as a small deviation 
    from the homentropic case, for which the tilt is negligible and 
    lines of constant $\delta\omega_z <0 $ are parallel to the vertical axis. 
    %    This implies that \emph{local} anti-cylonic motion, associated with $\delta\omega_z<0$, is horizontal. 
    Now consider baroclinity as a perturbation to this configuration.  
    
    The discussion in \S\ref{baroclinic_effects}, together with Fig. \ref{nonhomentropic_case_dvz}, 
    suggests that baroclinity gives rise to positive 
    azimuthal vorticity (as evident from the meridional flow pattern).
    We can produce azimuthal vorticity by tilting a 
    vertical column negative of $\dd\omega_z$ in the azimuthal direction indicated in 
    Fig. \ref{nonhomentropic_case_vort3d_pz}. By such a tilt, what was purely 
    horizontal \emph{local} anti-cyclonic motion, associated with $\dd\omega_z<0$ being a vertical column, now has a non-zero projection
    onto the meridional plane. This results in the meridional vortical motion demanded by the baroclinity in
    nonhomentropic flow. In other words, baroclinity has converted some of the horizontal motion of the 
    homentropic flow into vertical motion. 

%      If we consider baroclinity as a small 
%      perturbation to the homentropic solution, then a way to achieve
%      (positive) azimuthal vorticity, as
%      demanded by baroclinity, is a small tilt
%      shown in Fig. \ref{nonhomentropic_case_vort3d_pz}. 
      
\subsubsection{$m=5$}
\begin{figure}[!t]
  \centering
  \includegraphics[scale=.425,clip=true,trim=0cm 2cm 0cm
    0.2cm]{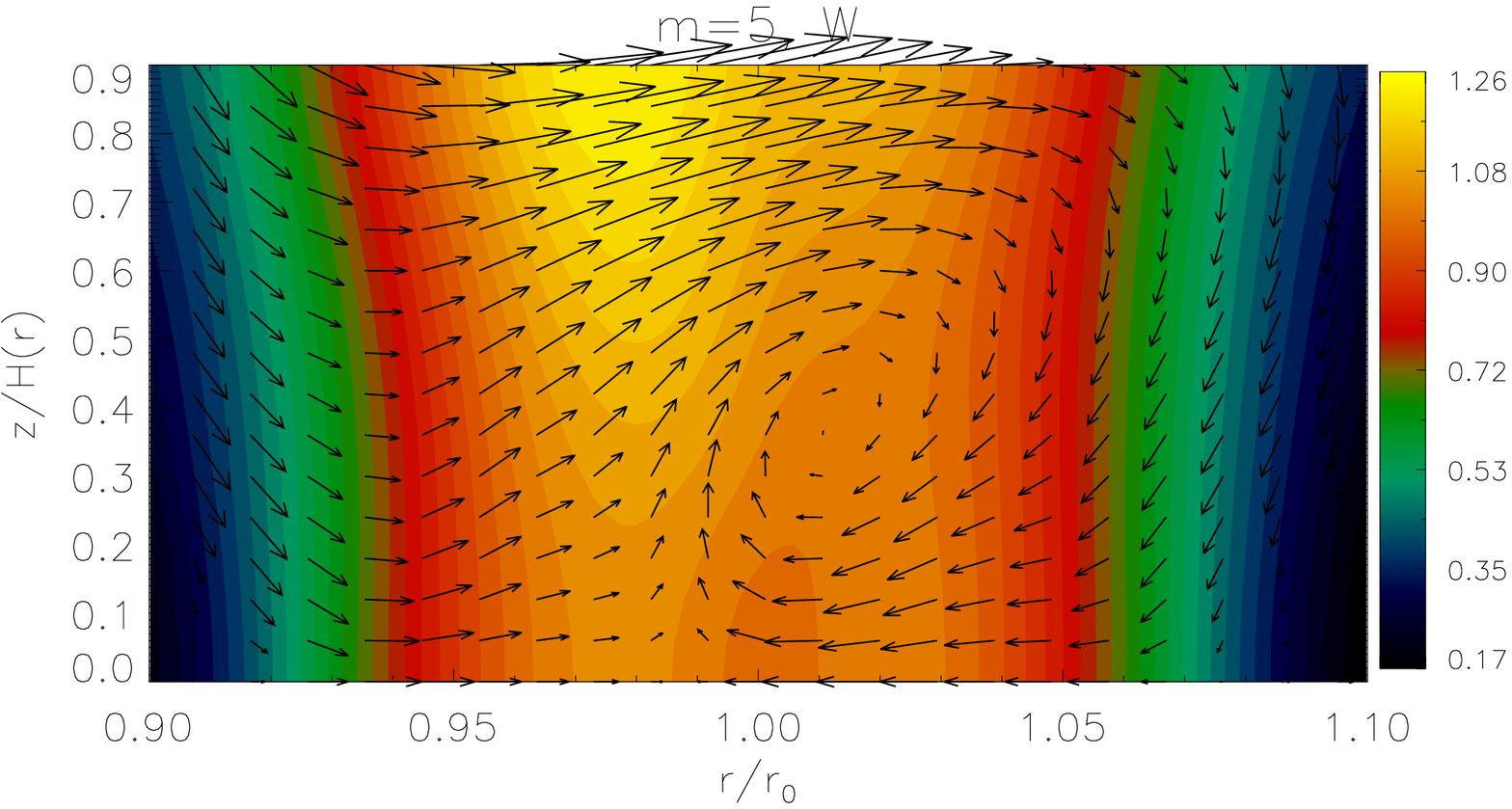}
  \includegraphics[scale=.425,clip=true,trim=0.0cm
    .48cm 0.cm 0.2cm]{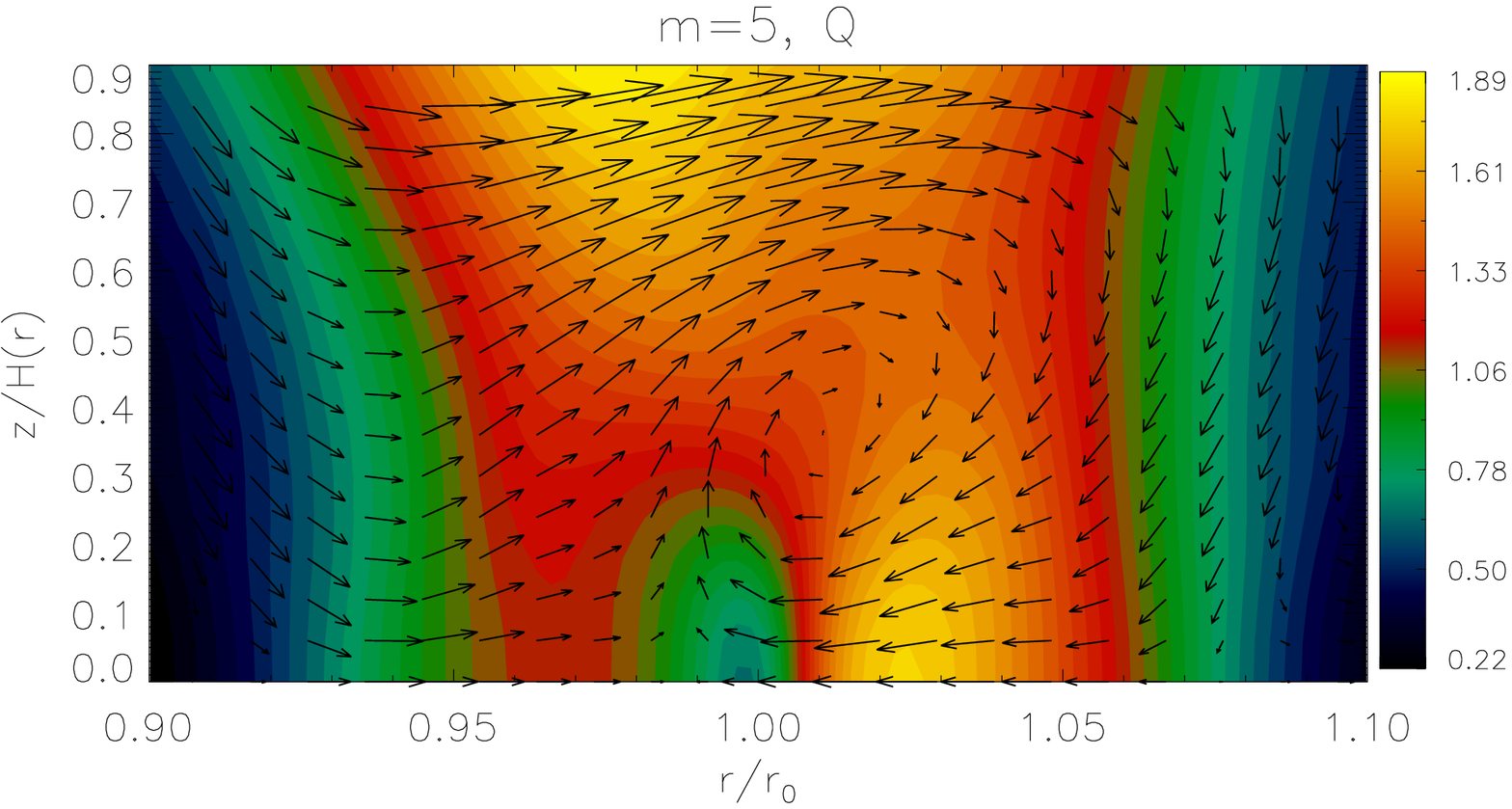}
  \caption{Pressure ($W$, top) and density ($Q$, bottom) perturbation
    for the $m=5$ mode in the nonhomentropic case 3a. The meridional
    flow is also shown. 
  \label{nonhomentropic_m5}}
\end{figure}

%HR similar result, slender disk similar result 
The meridional flow varies with $m$. Fig. \ref{nonhomentropic_m5} shows
the $m=5$ solution for the 
setup of case 3a. We focus on the region $R\in[0.9,1.1]r_0$ because
higher-$m$ modes are not as well-localized as low-$m$
\citep{lin11a}. %unless strongly unstable %solution %may be influenced by simple bc  
It displays stronger vortical motion than
the fiducial run with $m=3$, even though the growth rates are similar
($\nu/m\Omega_0=0.1051$ for $m=5$). The pressure and density
perturbations have noticeable vertical structure, with $W$ typically
increasing away from the midplane. This qualitatively differs from
homentropic cases. 

\subsection{Solid upper boundary}
In the above example, it is perhaps not surprising that entropy
perturbations became more negative away from the midplane, because the free
boundary condition demands $|Q|>|W|$ at $Z=Z_s$.  

\begin{figure}[!ht]
  \centering
%  \includegraphics[scale=.425,clip=true,trim=0cm 0.cm 0cm
  %    0cm]{n1.5_gmma2.5_noverp_vel3d_rz_dvz_nofill.ps}
  \includegraphics[scale=.425,clip=true,trim=0cm 0.cm 0cm
    0cm]{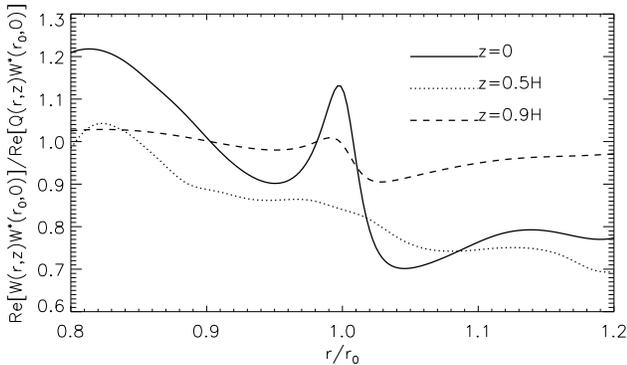}
  \caption{Ratio of pressure to density perturbations at $\phi_0$, for
    the nonhomentropic disk with solid upper disk boundaries
    (case 3b). A ratio above unity implies positive entropy 
    perturbation. 
    \label{novperp}}
\end{figure}

We have re-calculated this mode with a solid upper disk boundary
(case 3b). Numerically, this condition forces $W\simeq Q$ at 
$Z=Z_s$. Fig. \ref{novperp} shows the ratio of pressure to 
density perturbation. The entropy perturbation at
intermediate heights is still typically negative, suggesting this to
be an intrinsic feature of the instability in these disk models. The
flow pattern is very similar to case 3a.  

%survey across different gammas
\subsection{Effect of $\gamma$ on vertical flow}
\begin{figure}[!ht]
  \centering
  \includegraphics[scale=.425,clip=true,trim=0cm 0.cm 0cm
    0cm]{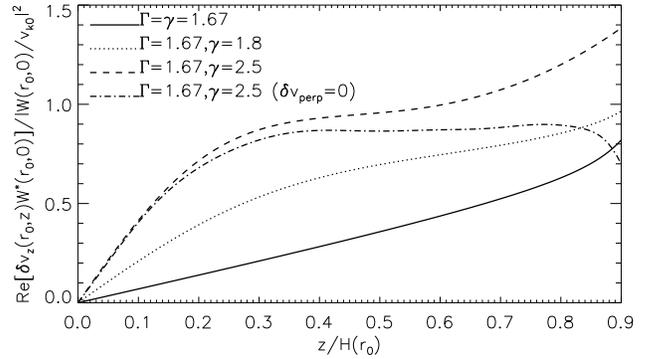}
  \caption{Normalized vertical velocities at the vortex core
    $(r_0,\phi_0)$ as function of $z$, for several values of $\gamma$
    with fixed $\Gamma=1.67$. The dash-dot line employed a solid upper
    disk boundary, other cases use the free boundary condition. 
    \label{compare_dvz}}
\end{figure}

When $\gamma\neq\Gamma$, the presence of buoyancy forces is expected
to modify the vertical flow associated with the RWI. In
Fig. \ref{compare_dvz} we compare the vertical velocity at the
vortex core for a range of $\gamma$. 
%Although we are
%effectively comparing gas of different thermodynamic properties, this
%set of runs is useful because the disks have indentical background
%density, pressure and velocity profiles.  

As we increase $\gamma$, the magnitude of vertical flow
increases, with an increasingly complicated $z$-dependence. In the 
homentropic case ($\gamma=1.67=\Gamma$), $\delta v_z $ is essentially linear in
$z$, consistent with the analytical models of \cite{umurhan10}.  
For $\gamma=2.5>\Gamma$, near the midplane $\delta v_z$ is still linear in
$z$, but away from $z=0$ the increase in $\dd v_z$ starts to level
off at $z=0.3H$ due to the development of meridional vortical motion. The leveling off 
occurs for both types of upper disk boundary conditions. 
%for both types of upper disk boundary;conditions; associated with the development of
%meridional vortical motion away from the midplane.
This results in a `step' in the case of a
solid upper boundary (centered about $z=0.5H$), but for the free upper boundary 
$\dd v_z$ increases again at large $z$. Since the RWI
is a global instability in the vertical direction, vertical
boundary conditions can affect the flow throughout the fluid 
column, though the extent of which depends on the equation of state
\citep[][see also \S\ref{isothermal}]{lin12c}. 
%% On the other hand, one physical origin for the
%% `leveling off' of $\delta v_z$ is the increasingly stable vertical
%% stratification away from the midplane.

 Let us examine the different contributions to vertical motion at
  co-rotation. At $(r_0,\phi_0)$, the vertical velocity is roughly 
\begin{align}\label{core_dvz}
  %  \dd v_z = \frac{\imgi}{\sbar H}\left[\frac{\p W}{\p Z} -
  %    \frac{2nZ}{(1-Z^2)}\left(W - \frac{\Gamma}{\gamma}Q\right)\right].  
  \dd v_z \sim -\frac{1}{\nu H}\left[\smash{\underbrace{\frac{\p W}{\p
          Z}}_\dagger} \!\!\!\phantom{\frac{1}{1}} +\phantom{\frac{1}{1}}\!\!\!
    \smash{\underbrace{\frac{2nZ\Gamma}{\gamma(1-Z^2)}\left(Q
      -\frac{\gamma}{\Gamma}W\right)}_\ddagger}\right].  \\ \notag
\end{align}

This equation is obtained from Eq. \ref{vz} by evaluating it 
at co-rotation radius (where
$\sbar\simeq -\ii\nu$), and inserting expressions for the
pressure length-scale for polytropic backgrounds. 
The first term ($\dagger$) represent pressure forces and is present
for all values of $\gamma$. For homentropic flow, ($\dagger$) is
the only source of vertical motion, and in this case $W$ decreases
with height. The second term ($\ddagger$) is only present if $\gamma\neq\Gamma$. 
Recall the quantity $Q-\gamma W/\Gamma=\bar{S}$ defined in
\S\ref{baroclinic_effects}, where it appeared as a baroclinic source term and 
we argued $\bar{S}\leq0$ at the vortex core (see also Fig. \ref{Sbar}). Then at the vortex core,
($\ddagger$) contributes positively to $\dd v_z$ along the vertical
direction, but vanishes at endpoints.     

In the nonhomentropic example (case 3a, $\gamma=2.5$) the
function $W$ increases with height at the vortex core (Fig. \ref{nonhomentropic_case}),
implying ($\dagger$) contributes negatively to $\dd v_z$.  
  The contribution from ($\ddagger$) and $(\dagger)$ have opposite
  signs, but the fact
that we observe positive vertical velocity shows that ($\ddagger$) is 
typically larger in magnitude than ($\dagger$). That is,
  baroclinity typically outweigh vertical pressure gradients.
  
\subsubsection{The role of $N_z^2\neq 0$}\label{nz_nonzero}
Notice even when $\gamma$ is only slightly larger than $\Gamma$,
the vortex core vertical velocity is quite different from the homentropic case 
(i.e. case 1 with $\gamma/\Gamma=1.08$ in Fig. \ref{compare_dvz}). To see the role of entropy gradients, 
or equivalently the effect of non-zero buoyancy frequency, we follow \cite{kato01} and 
make the following approximations. For generality, below 
we shall not specialize to a polytropic background. 

Consider a height at which $\ihs\gg\ils$, which is generally
true away from the midplane of a thin disk. Furthermore, suppose radial 
velocities are not much larger than vertical velocities in the region of interest (co-rotation). 
Then we can neglect the $\dd v_r$ term in the linearized energy equation, and eliminate $Q$ between 
Eq. \ref{vz} and Eq. \ref{lin_energy} to obtain,
\begin{align}\label{vz_approx} 
%\dd v_z \simeq \frac{\imgi}{\sbar}\left[\frac{\p W}{\p z} + \left(\frac{\p\ln{\rho}}{\p z} - \frac{1}{H_p}\right)\right]
%+\frac{N_z^2}{\sbar^2}\dd v_z,
\dd v_z \simeq -\frac{1}{\nu}\left[\frac{\p W}{\p z} + \left(\frac{\p\ln{\rho}}{\p z} - \frac{1}{H_p}\right)W\right]
-\frac{N_z^2}{\nu^2}\dd v_z,
\end{align}
which is \citeauthor{kato01}'s Eq. 21 evaluated at co-rotation. 
Because $\nu\ll\Omega_0$ for the modes considered and $N_z\sim \Omega$ away from the midplane, 
for nonhomentropic flow we have $N_z^2/\nu^2\gg1$ and should expect the balance
\begin{align}\label{kato_vz}
\dd v_z \sim  -\frac{\nu}{N_z^2}\frac{\p W}{\p z} 
 \underbrace{- \frac{\nu}{N_z^2}\left( \frac{\p\ln{\rho}}{\p z} 
- \frac{1}{H_p} \right)W}_{- \nu\rho\left(\frac{\p p}{\p z} \right)^{-1}W}, \quad N_z^2\neq0%\\
%&=  -\frac{\nu}{N_z^2}\frac{\p W}{\p z} - \nu\rho\left(\frac{\p p}{\p z} \right)^{-1}W \notag
\end{align}
near co-rotation radius.
The second term on the RHS is just buoyancy (and does not
explicitly depend on $\gamma$). This expression should be compared with
that for strictly homentropic flow, 

%% Comparing the coefficient of $\p_zW$ in this expression to that in
%% the strictly homentropic case, where 
\begin{align*}
  \dd v_z \sim -\frac{1}{\nu}\frac{\p W}{\p z}, \quad N_z^2\equiv 0. 
\end{align*}

We see that for $N_z\equiv0$, pressure gradients are entirely
responsible for vertical flow, whereas for $N_z\neq0$, $\dd v_z$ is result of
a combination of pressure and buoyancy forces. The importance of
pressure gradients also differ, because the coefficients of $\p_zW$ are different
in each case (by a factor $\nu^2/N_z^2$). 
Furthermore, the ratio of the first to second term in
Eq. \ref{kato_vz} is approximately 
\begin{align}
  \frac{(\nu/N_z^2)\p W/\p z}{\nu\rho(\p p/\p z)^{-1}W}\sim
  \frac{\Omega^2}{N_z^2}\frac{\p\ln{W}}{\p\ln{z}}.
\end{align}
Since $N_z$ increases with height, far away from the midplane we
expect buoyancy forces to dominate in the nonhomentropic case. 

We
conclude that the origin of vertical motion at co-rotation is qualitatively different
between homentropic and nonhomentropic flow (especially away 
from the midplane), as suggested by numerical results in the
previous section.

\subsection{Fixed $\gamma$, variable $\Gamma$}\label{varpolyn}
We now fix the adiabatic index to $\gamma=1.4$, as is typical for 
accretion disk models. Then we require $n>2.5$ for axisymmetric stability. 
With other parameters fixed, increasing $n$ would decrease the bump in 
disk thickness and reduce growth rates \citepalias{lin12}. 
To avoid potential numerical issues associated with small $|\sbar|$ at
co-rotation, we adopt $h=0.2$ for cases 5---8, so that growth rates
remain $O(0.1\Omega_0)$.

Table \ref{linsims} shows that by setting $\gamma\neq\Gamma=1.33$ (case 6), 
thereby introducing entropy gradients, $\avg{\theta_m}$ has increased 
from the homentropic case 5. This is consistent with the trend in 
cases 0---4.    

\citetalias{lin12} found that when $n$ is increased but other
parameters fixed,  the flow at the vortex core became less
three-dimensional. For cases 5---8, we find the average value of  
$\theta_m$, when taken over $R\in[0.98,1.02]r_0$, is 
$0.46,\, 0.63,\, 0.61$ and $0.56$ for $n=2.5,\,3.0,\,3.5$ and $4.0$,
respectively. The flow at $(r_0,\phi_0)$ in fact becomes more
three-dimensional when it is nonhomentropic although $n$ has
increased (case 5 $\to$ case 6). 

The small decrease in the above values of three-dimensionality at
$(r_0,\phi_0)$ from $n=3.0$ to $n=4.0$ is likely related to increased 
radial flow across $r_0$ associated with vortical motion in the 
$(r,z)$ plane. Cases 6---8 display similar dependence of $\dd v_z$ on
$z$ as the nonhomentropic example (case 3a, see Fig. \ref{compare_dvz}).     

%It appears that the competing effect of
%reducing in $|\dd v_z|$ due to increased $n$ and increasing $|\dd
%v_z|$ due to increased $|\gamma-\Gamma|$ (and growth rate), 
%core vz
%3d-ness still goes up with

%% \begin{figure}[!ht]
%%   \centering
%%   \includegraphics[scale=.425,clip=true,trim=0cm 0.cm 0cm
%%     0cm]{compare_dvz_varpolyn}
%%   \caption{Normalized vertical velocities at the vortex core
%%     $(r_0,\phi_0)$ as function of $z$, for cases with fixed adiabatic
%%     index but variable polytropic index. 
%%     \label{compare_dvz_varpolyn}}
%% \end{figure}

\section{Isothermal limit}\label{isothermal}
We now examine the limit $\Gamma\to 1$, where the unperturbed disk becomes isothermal
, but perturbations are evolved with an adiabatic index
$\gamma=1.4$. We consider a nearly-isothermal polytropic background
and strictly isothermal backgrounds. These cases are treated
separately because the equilibrium structures have different
functional forms. A comparison between them provide another
check on our numerical results. 

\subsection{Large polytropic index}
We first consider setting $n=10$ to produce an almost radially
isothermal equilibrium with $p\propto \rho^{1.1}$. This allows us to use the
numerical code as set up for polytropic equilibria without
modification. We also adopt $A=2.5$ and $h=0.25$ for reasons given
in \S\ref{varpolyn}. The relatively large aspect-ratio does not
violate the thin-disk approximation as large $n$ implies the 
density decays rapidly away from the midplane. Also because of this,
we set the upper disk boundary at $Z_s=0.6$ to avoid very low 
densities. %All other parameters are the same as before. 

%3d-ness of core is 0.55
%3d-ness averaged over 0.9, 1.1 is 0.37
%%  Despite the larger
%% value of $\gamma/\Gamma$ than case 8 (see Table \ref{linsims}), the flow
%% three-dimensinality is the same.   

\begin{figure}[!t]
  \centering
  %% \includegraphics[scale=.425,clip=true,trim=0cm 1.cm 0cm
  %%   0cm]{n10.0_gmma1.4_wq_3d_real_zslices.ps}
  \includegraphics[scale=.425,clip=true,trim=0cm 1.83cm 0cm
    0.25cm]{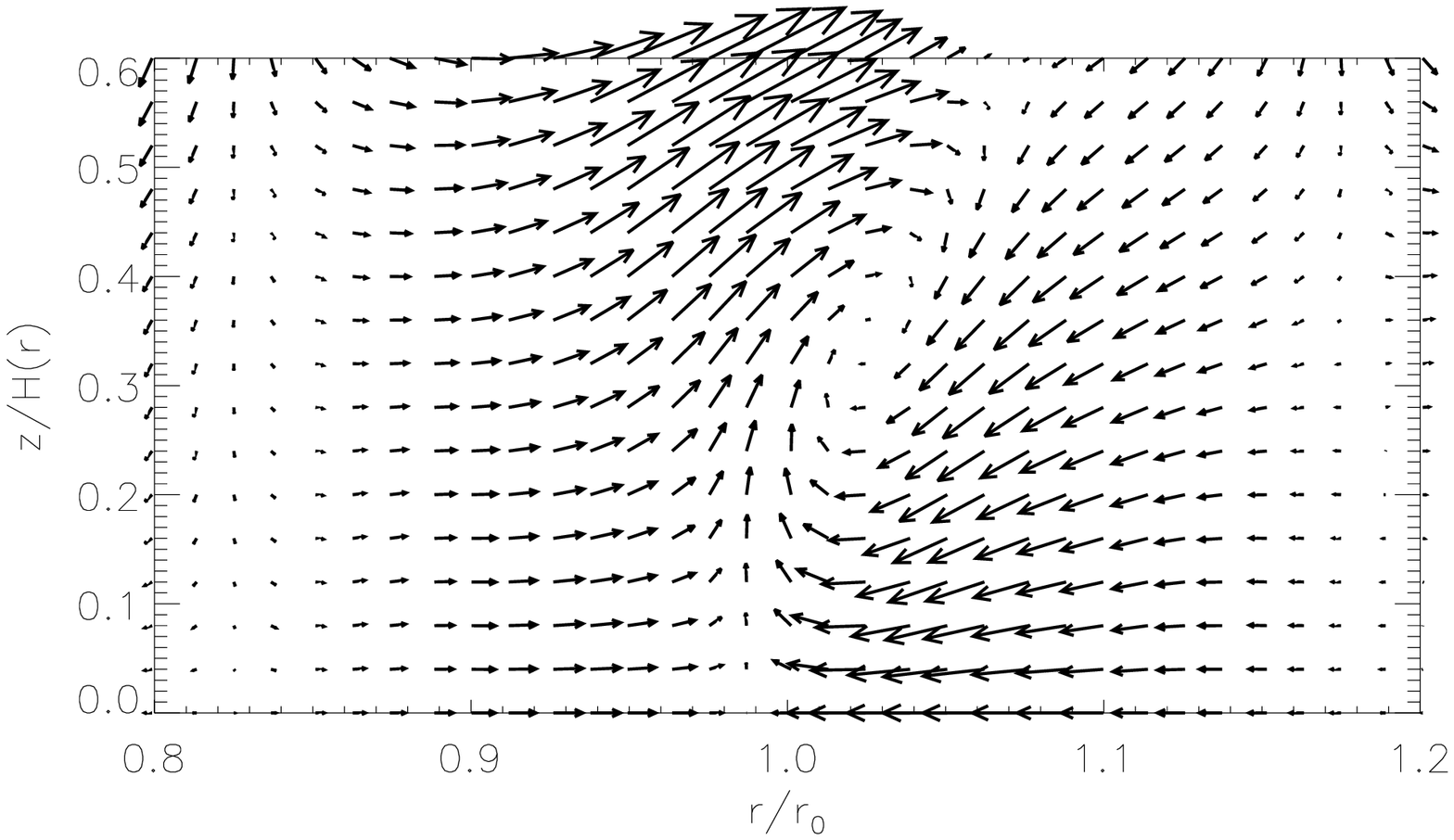}
   \includegraphics[scale=.425,clip=true,trim=0cm 0cm 0cm
    0.25cm]{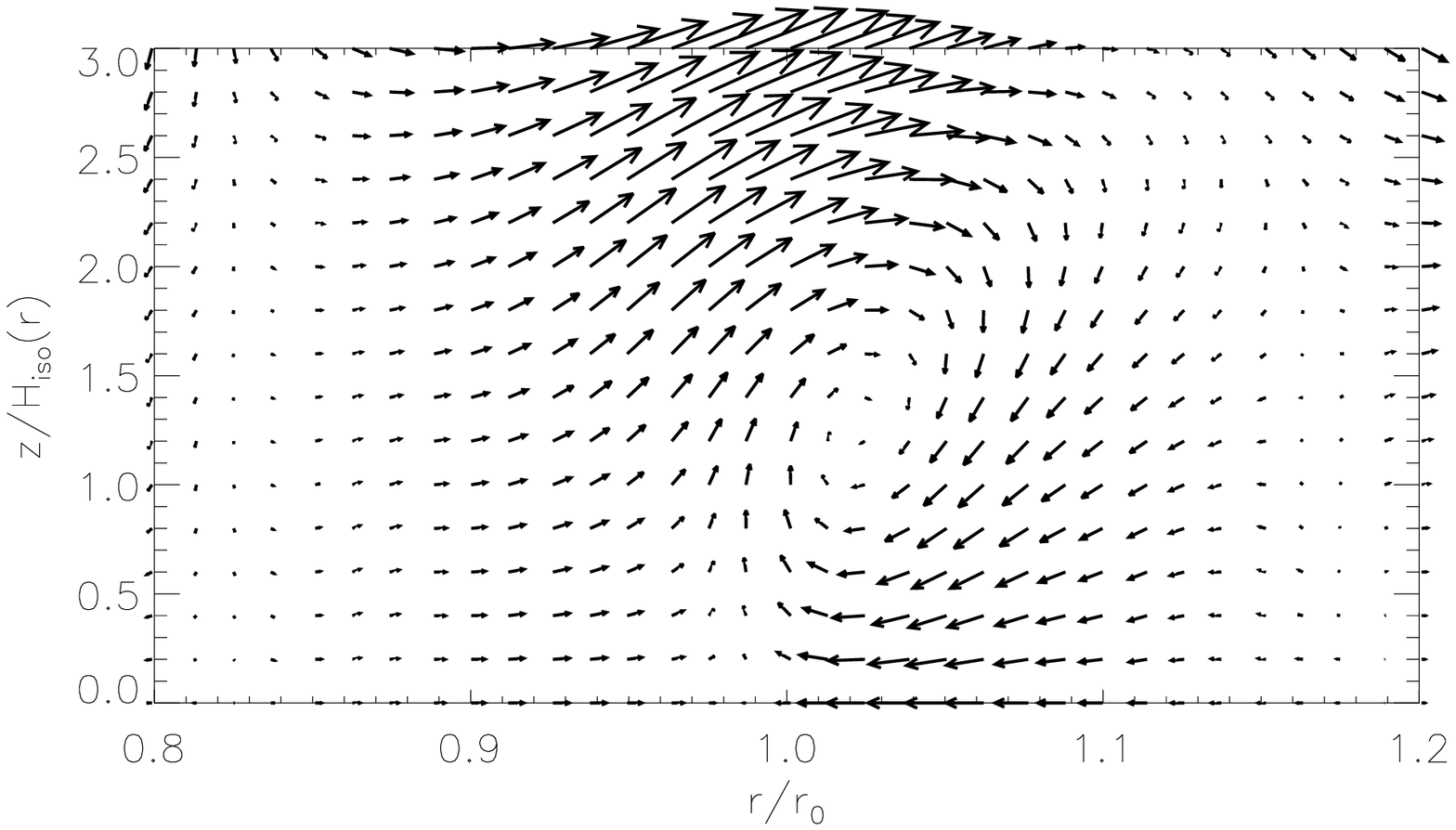}
  \caption{Perturbed meridional flow at $\phi=\phi_0$ for a $n=10$
    polytropic disk equilibrium (top) and a strictly isothermal
    equilibrium (bottom). 
    \label{polyn10}}
\end{figure}

For this setup we obtained $\omega/m\Omega_0=0.9883$,
$\nu/\Omega_0=0.1375$ and $\avg{\theta_m}=0.35$. The top panel of
Fig. \ref{polyn10} shows the meridional flow at the vortex core. The
vortical motion is distinct and more apparent than case 3a, despite
the smaller value of $\gamma/\Gamma$ in the present case. 
However, apart from this difference, the solution is qualitatively
similar to case 3a.   

\subsection{Strictly isothermal equilibrium}
%Because we have formulated the linear problem in terms of entropy and
%pressure length-scales, in order to calculate the RWI for strictly
%isothermal disks, we can just replace 
Modifications to our standard setup are required to treat  
disk equilibria with $p=\ciso^2\rho$ ($\Gamma\equiv 1$), where 
the constant sound speed $\ciso=\Hiso\Omega_k$, $\Hiso=\hiso
r_0(r/r_0)^{3/2}$ is the isothermal scale-height, and $\hiso$ is the
characteristic aspect-ratio at $r_0$. 
%We set $\ciso=\Hiso
%\Omega_k$, where $\Hiso\propto r^{3/2}$ is the isothermal scale-height
%and we define $\hiso=\Hiso/r_0$. 
The dimensionless vertical 
co-ordinate is now $Z=z/\Hiso$. The isothermal atmosphere is
exponential, $g(Z)=\exp{(-Z^2/2)}$, so there is no surface. In
practice we choose a finite vertical domain, i.e. $Z=Z_s$ represents a
constant number of isothermal scale-heights above the midplane. 

In the linear code we simply replace expressions for the entropy
and pressure length-scales by those corresponding to the isothermal
disk: the function $H\to\Hiso$ and $g(Z)$ becomes the Gaussian above. 
%% We set parameter values to mimic the large-$n$ polytrope in the 
%% previous section:  $Z_s=3$ and $\hiso=0.05$,
%% so that for both setups the pressure is reduced by roughly the 
%% same factor in going from the midplane to the upper boundary, and the
%% midplane temperature is roughly equal at $r_0$. 
We choose $Z_s=3$ and $\hiso=0.05$, so the isothermal disk has roughly
the same temperature as that in the midplane of the large-$n$ 
polytrope considered above (at $r_0$). In going from the
midplane to the upper boundary, the density is also reduced by
approximately the same factor for both cases.

\begin{figure}[!t]
  \centering
  \includegraphics[scale=.425,clip=true,trim=0cm 0.cm 0cm
    0cm]{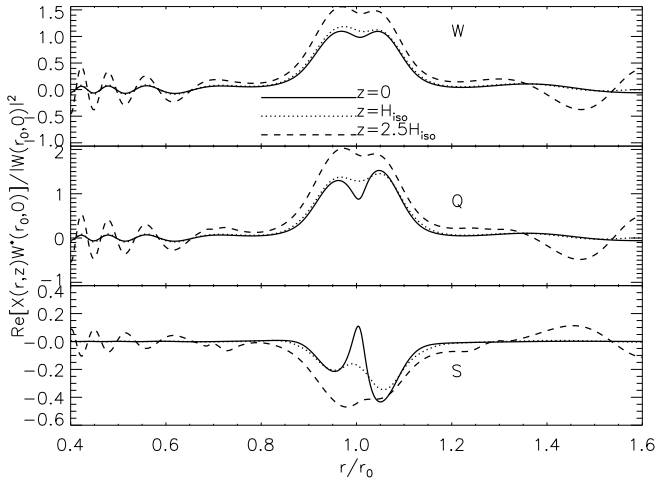}
  \caption{Pressure (top, $W$), density (middle, $Q$) and entropy
    (bottom, $S$) for a globally isothermal background. 
    \label{isothermal_case}}
\end{figure}

We obtain $\omega/m\Omega_0=0.9860$, $\nu/\Omega_0=0.1008$ and
$\avg{\theta_m}=0.39$. The perturbations plotted in
Fig. \ref{isothermal_case} are similar to case 3a, so we expect these
are features of the RWI in nonhomentropic flow, rather than
associated with the chosen parameter values. The perturbed meridional
flow shown in Fig. \ref{polyn10} (bottom panel) is in qualitative
agreement with the large-$n$ polytrope. 
The result is, however, quite different to
isothermal linear perturbations, for which \cite{meheut12} found the 
vertical velocity appears to have a node at $r_0$ (see their Fig. 3d where the vertical velocity 
changes sign across co-rotation radius,  
i.e. the fluid column is hydrostatic there). Here, there is clearly vertical motion at co-rotation. 
Note that both $\gamma/\Gamma$
and the growth rate are slightly smaller than the nonhomentropic case 3a,
but here the vortical motion is more prominent. 
%contribution from higher m?

\begin{figure}[!t]
    \centering
    \includegraphics[scale=.425,clip=true,trim=0cm 0cm 0cm
      0.25cm]{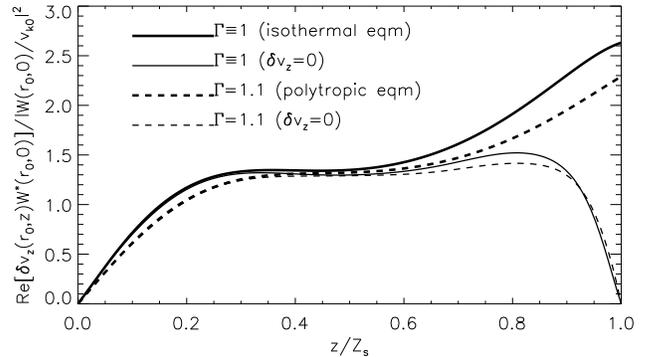}
    \caption{
        Vertical velocity as a function of $z$ at the vortex core
        $(r_0,\phi_0)$, for the $n=10$ 
        polytropic disk equilibrium (dashed) and a strictly isothermal
        equilibrium (solid) shown in 
        Fig. \ref{polyn10}, with free upper boundaries (thick lines). Corresponding thin 
        lines impose zero vertical velocity at $z=Z_s$ (growth rates increased by less than $0.5\%$ from the free boundary condition).
        Notice that changing upper disk boundary conditions only 
        affected the solution near $z=Z_s$ (cf. Fig. \ref{compare_dvz}). 
        This is consistent with \cite{lin12c}, who found the
        influence of upper disk boundary condition to diminish with
        increasing polytropic index $n$. 
      \label{polyn10_vz}}
\end{figure}

  Fig. \ref{polyn10_vz} shows the vertical 
  velocity at the vortex core as a function of height. The strictly
  isothermal background (thick solid) has a slightly larger $\dd v_z$ than the
  large-$n$ polytrope (thick dashed). This is consistent 
  with previous findings that vertical motions oppose the RWI
  \citep{lin12c}, as the former case has a smaller growth rate than
  the latter. The thick lines are qualitatively similar to case
  3a in Fig. \ref{compare_dvz}, but these are not directly comparable 
  because the present case differs in both the background structure and adiabatic index  to 
  those in Fig. \ref{compare_dvz}. 
  
\begin{figure}[!t]
  \centering
  \includegraphics[scale=.425,clip=true,trim=0cm 0cm 0cm
    0.25cm]{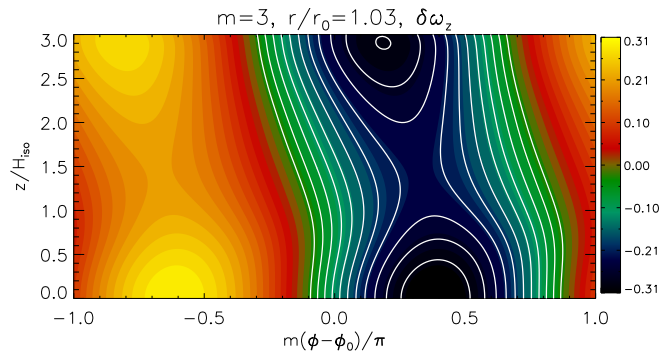}
  \caption{
      Vertical vorticity perturbation, $\delta\omega_z$, in the
      $(\phi,z)$ plane at $r=1.03r_0$ for the strictly isothermal
      background. Regions of $\delta\omega\leq0$ are delineated by
      while lines. The center of meridional vortical motion identified
      in Fig. \ref{polyn10} occurs at height $z\sim
      H_\mathrm{iso}$.  The azimuthal range
        $\phi-\phi_0\in[-0.5,0.5]\pi/m$ corresponds to anti-cyclonic
        motion about the vortex core.
        [ A plot for $r=1.02r_0$ also display
      tilted lines of constant $\delta\omega_z$, but in that case $\delta\omega_z > 0$
      at $(\phi_0, H_\mathrm{iso})$.]
    \label{iso_vort3d_pz}}
\end{figure}

  We illustrate again a correlation 
  between meridional vortical flow and a tilted column of negative
  vertical vorticity perturbation in Fig. \ref{iso_vort3d_pz}. The figure is
  qualitatively similar to that for polytropic backgrounds (case 3a in
  Fig. \ref{nonhomentropic_case_vort3d_pz}).  We find an average tilt
    of $1-~\avg{\cos{\theta}}_Z=0.0084
    \ll1$, so the vorticity column is nearly vertical. 
  
  %% but the tilt magnitude
  %%   is larger for the isothermal background. 
  %Note that we chose
  %$r=1.03r_0$ as the radial slice, so that $\delta\omega_z <0$ along
  %the vertical line $(\phi_0,z)$. 
  %%  A plot for $r=1.02r_0$ also display
  %% the tilt, but in that case $\delta\omega_z$ was found to be positive
  %% at   

\subsection{A nonlinear simulation}
%boundary conditions
%compare Q at different heights
%compare W in rz plane - vortical motion
We have also performed global 3D hydrodynamic simulations
using the \zeus  finite-difference code \citep{hayes06}. As the focus
of this work is the linear problem, though, we defer a full discussion
of these nonlinear simulations to a follow-up paper. Our priority here
is to verify the  vortical motion in the meridional plane, which
appears characteristic in  the linear RWI solution for  nonhomentropic
flow.   

\subsubsection{Setup}
We use spherical polar co-ordinates $(\rsph,\theta,\phi)$ to describe
the disk, taken to be initially strictly isothermal as described above.  The
computational domain is $\rsph\in[0.2,2.0]r_0$,
$\theta\in[\theta_\mathrm{min},\pi/2]$, $\phi\in[0,2\pi]$ and is
divided into $(512,48,512)$ zones, with
$\tan{(\pi/2-\theta_\mathrm{min})}=3\hiso$ and $r_0=10$. 
The grid is logarithmically spaced in radius and uniformly spaced in 
the angular co-ordinates. Boundary conditions 
are outflow in $\rsph$, reflection in $\theta$ and periodic in $\phi$.  
Additional damping to meridional velocities near radial boundaries are
employed to reduce reflections \citep{valborro07}.  

%The disk is initially strictly isothermal as described in the previous section, and the vertical domain is such that 
%$\tan{(\pi/2-\theta_\mathrm{min})}=3\hiso$. 

After some experimentation, we found it was most convenient to start with a smooth disk.
In this case, a surface density $\Sigma\propto r^{-3/2}$, and tapered toward 
the inner boundary \citep[as used in][]{lin12b}. We introduce 
the density bump at $r=r_0$ via source terms in the mass, momentum and thermal energy equations, over a time-scale of $10P_0$, 
where $P_0 \equiv 2\pi/\Omega_k(r_0) $. %% is the Keplerian orbital period at $r_0$
This reduces numerical transients associated with initialization with a localized bump which has large radial gradients. 
%At $t=10P_0$ we add small-amplitude random
%radial velocity perturbations. 

We choose the bump amplitude $A=1.25$ 
and isothermal aspect-ratio $\hiso=0.1$, as employed by \cite{meheut12} so that we can check our results against theirs.  
We measure perturbations with respect to azimuthally averaged hydrodynamic quantities at $t=10P_0$.

\subsubsection{Results and comparison to linear flow}
\begin{figure}[!t]
  \centering
  \includegraphics[scale=0.62,clip=true,trim=0cm 0.cm 0cm
    0cm]{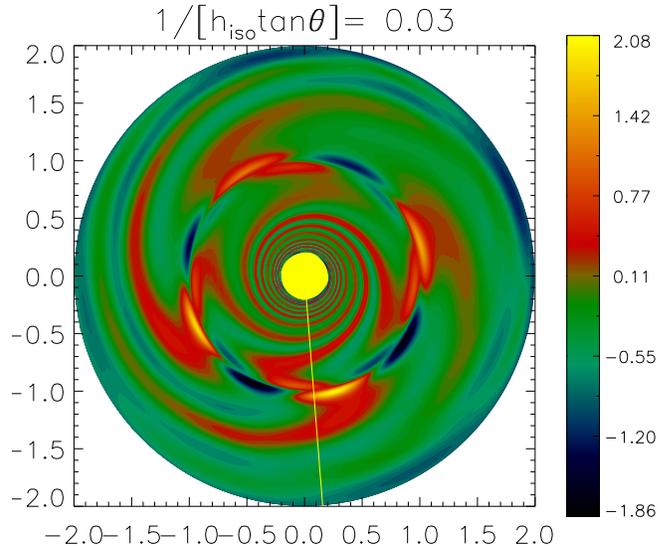}
  \caption{Nonlinear hydrodynamic simulation of the RWI in a
    nonhomentropic 3D disk, initially isothermal but
    evolved adiabatically. The axes are in units of $r_0$. The relative density perturbation 
    near the midplane, scaled by 100, is shown. This quantity is
    proportional to the $Q$ used in linear calculations.  The smallness of the density perturbation 
    implies that the snapshot corresponds to the linear phase of the instability. The drawn line   
    defines the vortex azimuth $\phi_0$ in Fig. \ref{lin_nonlin}---\ref{lin_nonlin_rz}.
    \label{hydro_polar_dens}}
\end{figure}

%complex freq. averaged over r_sph\in[0.8,1.2]r_0 is
%<growth rate>/omega0=      0.18870212
%<pattern speed>/m.omega0=      0.98988571
%
%complex freq. averaged over r_sph\in[0.9,1.1]r_0
%<growth rate>/omega0=      0.19579173
%<pattern speed>/m.omega0=      0.98709079
%theta (3dness) is 0.34
We focus on the earliest stage of the instability, when perturbation
amplitudes are small so comparison with linear calculations can be
made. Fig. \ref{hydro_polar_dens} shows the snapshot to be
examined,  taken at $t=23P_0$. A $m=4$ mode has developed 
  from numerical noise. Notice the double-peak in
density perturbation, which is also present in
Fig. \ref{isothermal_case}. Using the method described in Appendix
\ref{instant_growth}, we estimated the $m=4$ mode growth rate and
frequency to be $\nu/\Omega_0\simeq 0.194$ and %old value 0.189 did not include avg over theta
$\omega/m\Omega_0\simeq 0.990$, in agreement with \cite{meheut12}.
%consistent with linear calculations of %\cite{meheut12}.  
Although they assumed barotropic perturbations, 
whereas we simulate adiabatic evolution, our linear calculations
indicate growth rates are largely unaffected by entropy gradients (Table \ref{linsims}). 

\begin{figure}[!t]
  \centering
  \includegraphics[scale=.425,clip=true,trim=0cm 3.05cm 0cm
    1.3cm]{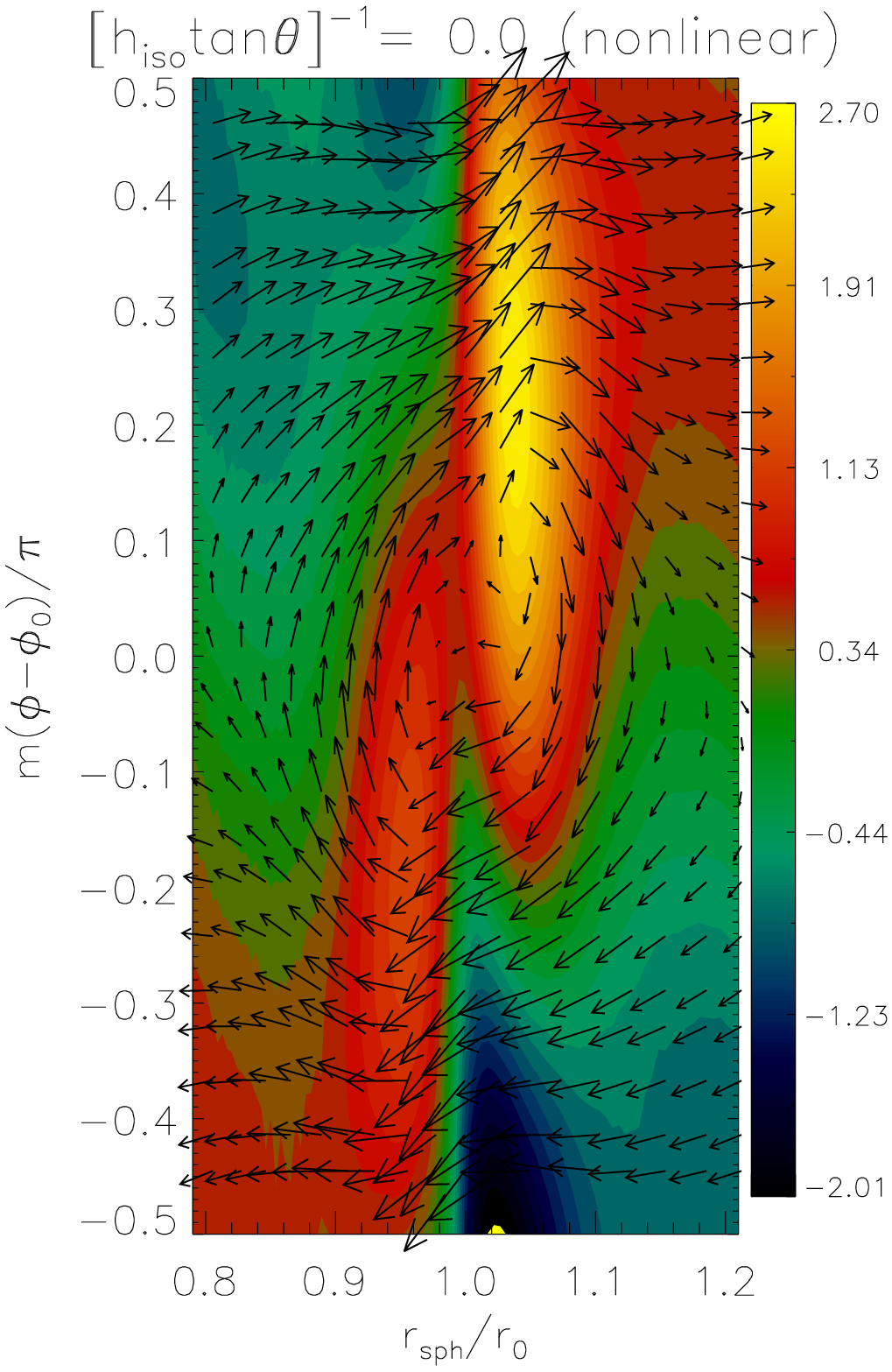}\includegraphics[scale=.425,clip=true,trim=2.2cm 
    3.05cm 0cm 1.3cm]{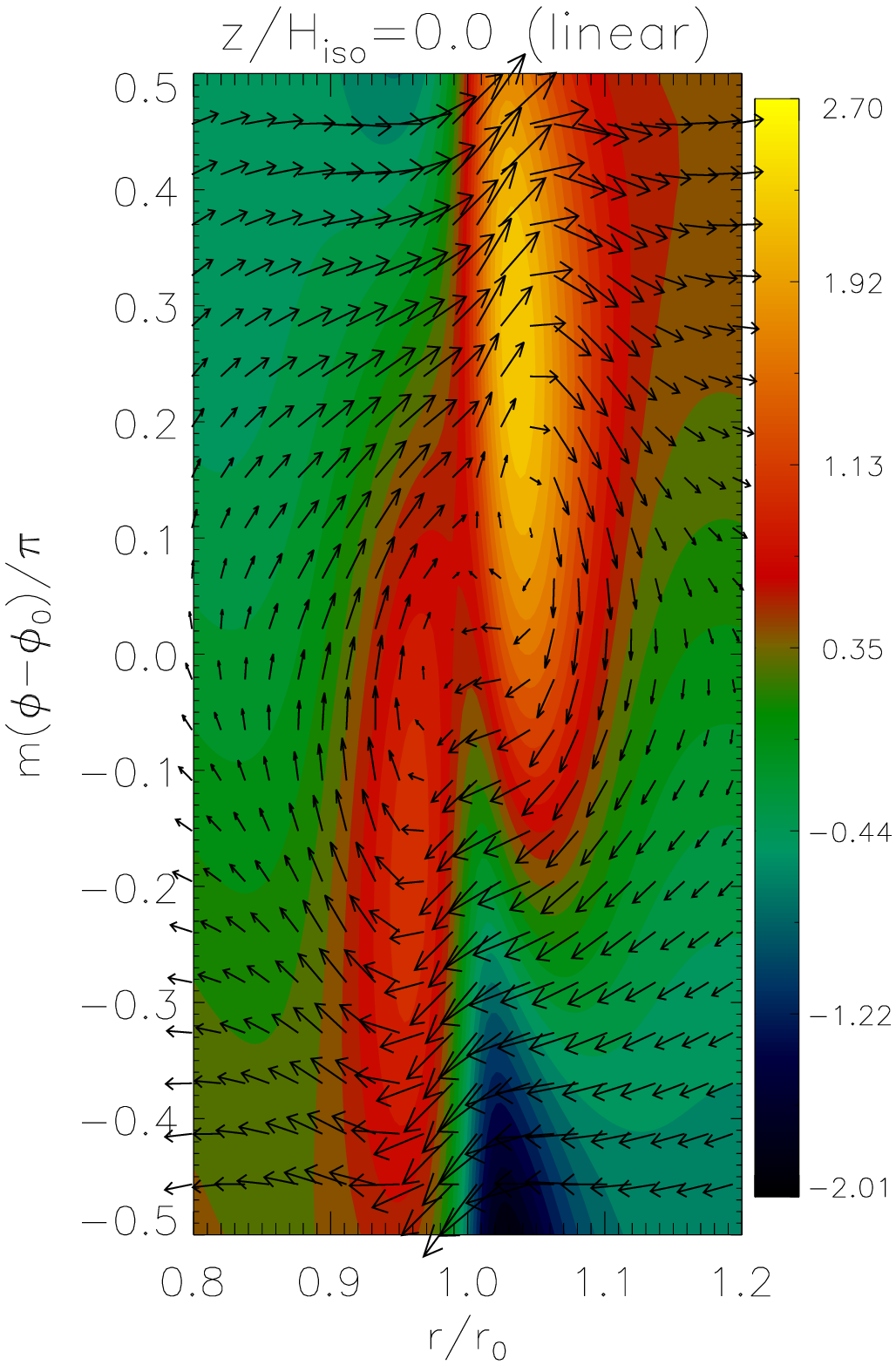}\\
  \includegraphics[scale=.425,clip=true,trim=0cm .0cm 0cm
    1.3cm]{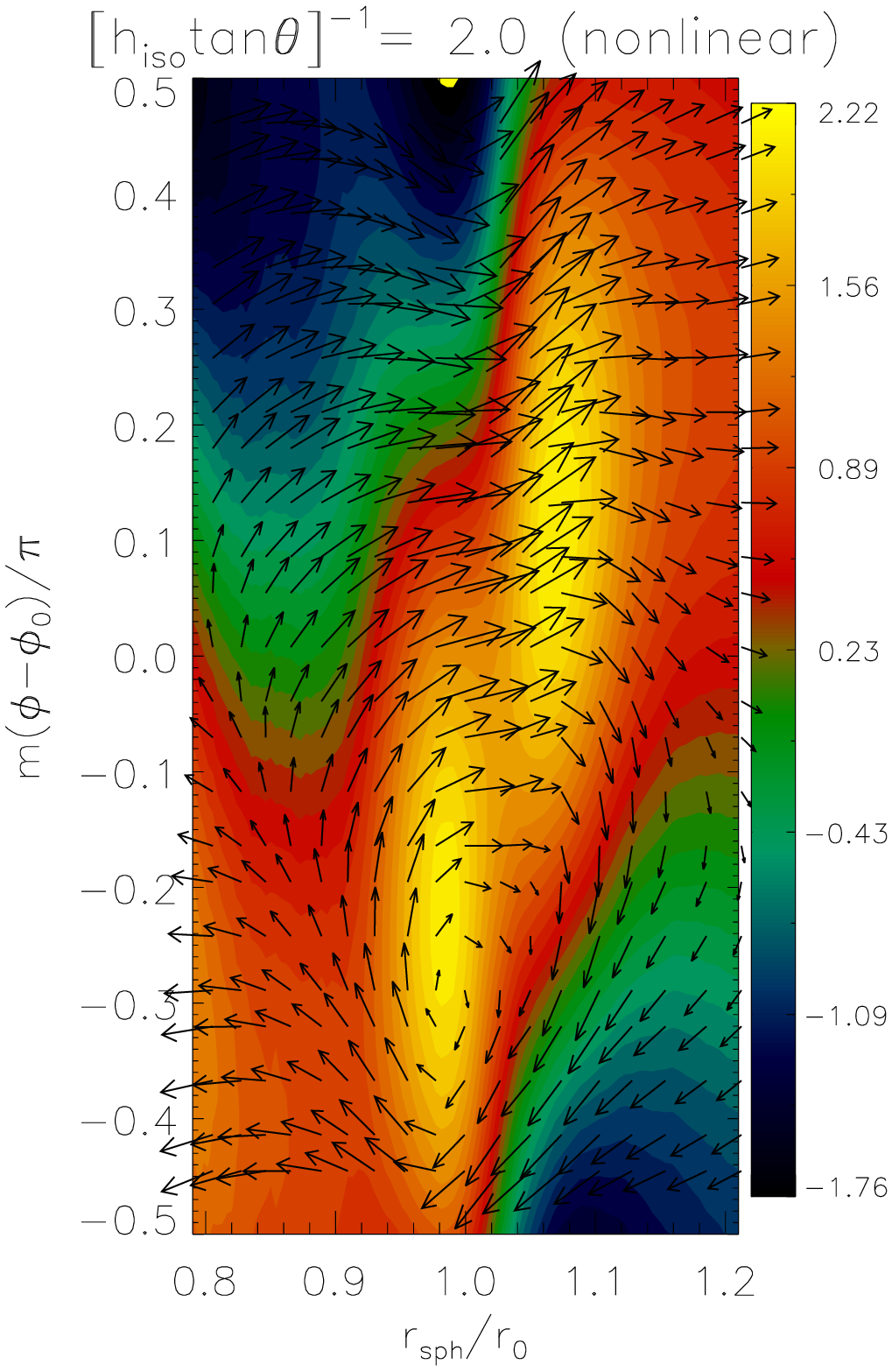}\includegraphics[scale=.425,clip=true,trim=2.2cm 
    .0cm 0cm 1.3cm]{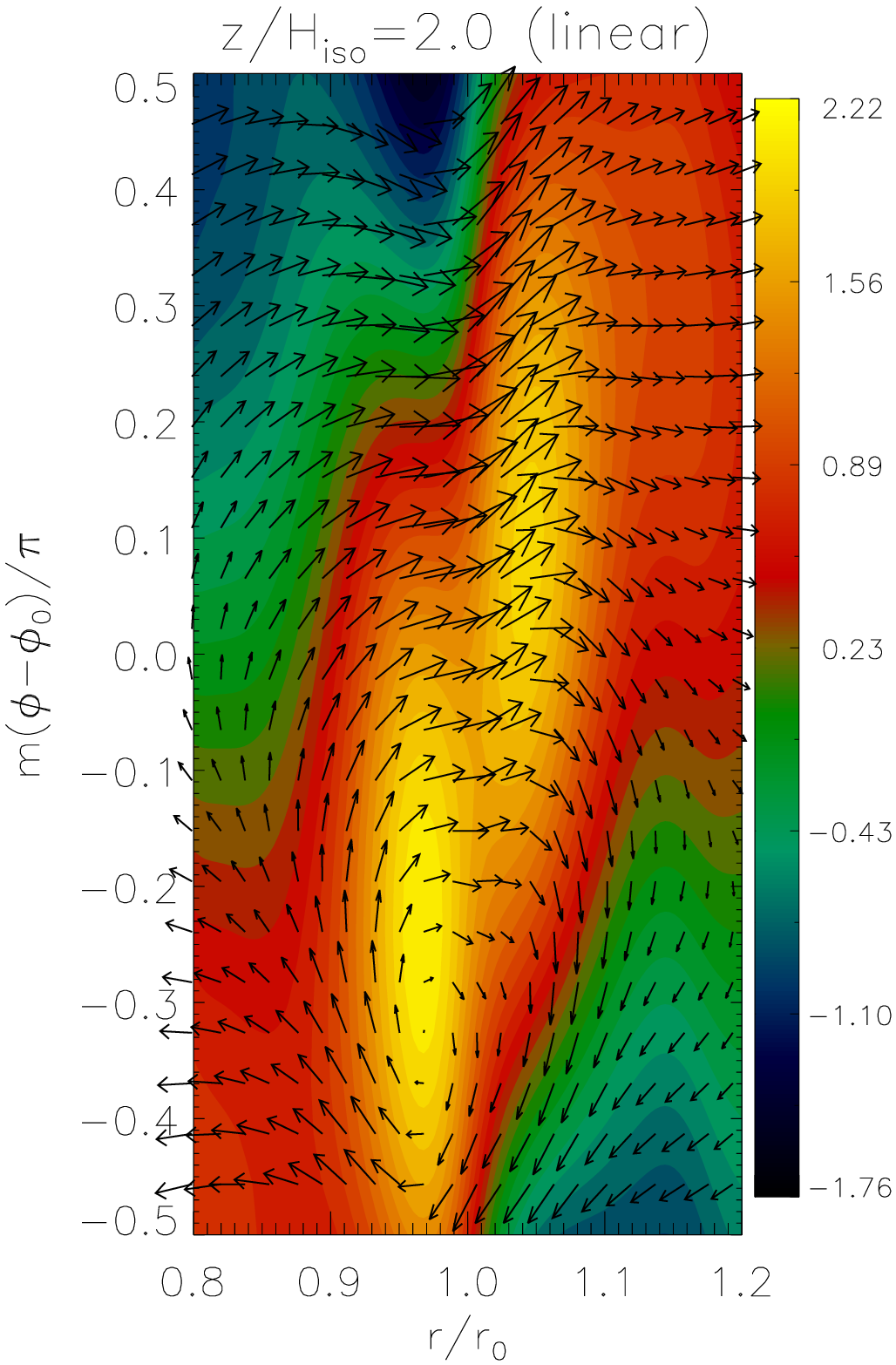}
  \caption{Normalized density perturbation, $Q$, associated with the
    RWI computed from a nonlinear hydrodynamic simulation (left) and
    the linear code (right), at the midplane (top) and at $2$
    scale-heights away from the midplane (bottom). The perturbed
    velocity field is also shown. The azimuthal wavenumber is $m=4$. 
  \label{lin_nonlin}}
\end{figure}

%use linear results for `big disk', theta_m is 0.39 
We have also computed this mode using the linear code as modified for
strictly isothermal equilibria, with a solid upper boundary. We obtain
growth rate and mode frequency $\nu/\Omega_0 = 0.1937$ and $\omega/m\Omega_0 = 
0.9896$, respectively. This is close to the nonlinear simulation. %\cite{meheut12} suggested
%smaller growth rate in the latter may be due to numerical viscosity. 
Fig. \ref{lin_nonlin} compares the density
perturbation $Q$ computed from the hydrodynamic simulation and linear code. They are broadly
consistent. The linear code also produces a bias toward the
over-density ahead of the vortex core at the midplane. Away from the
midplane, the center of the anti-cyclonic motion has shifted
downstream. 
This shows that, even within the linear regime, the vortex has
non-negligible vertical structure in the density perturbation (by comparing the two heights in Fig. \ref{lin_nonlin}). 
% for initially isothermal dis

\begin{figure}[!t]
  \centering
  \includegraphics[scale=.425,clip=true,trim=0cm 0.48cm 0cm
    0.4cm]{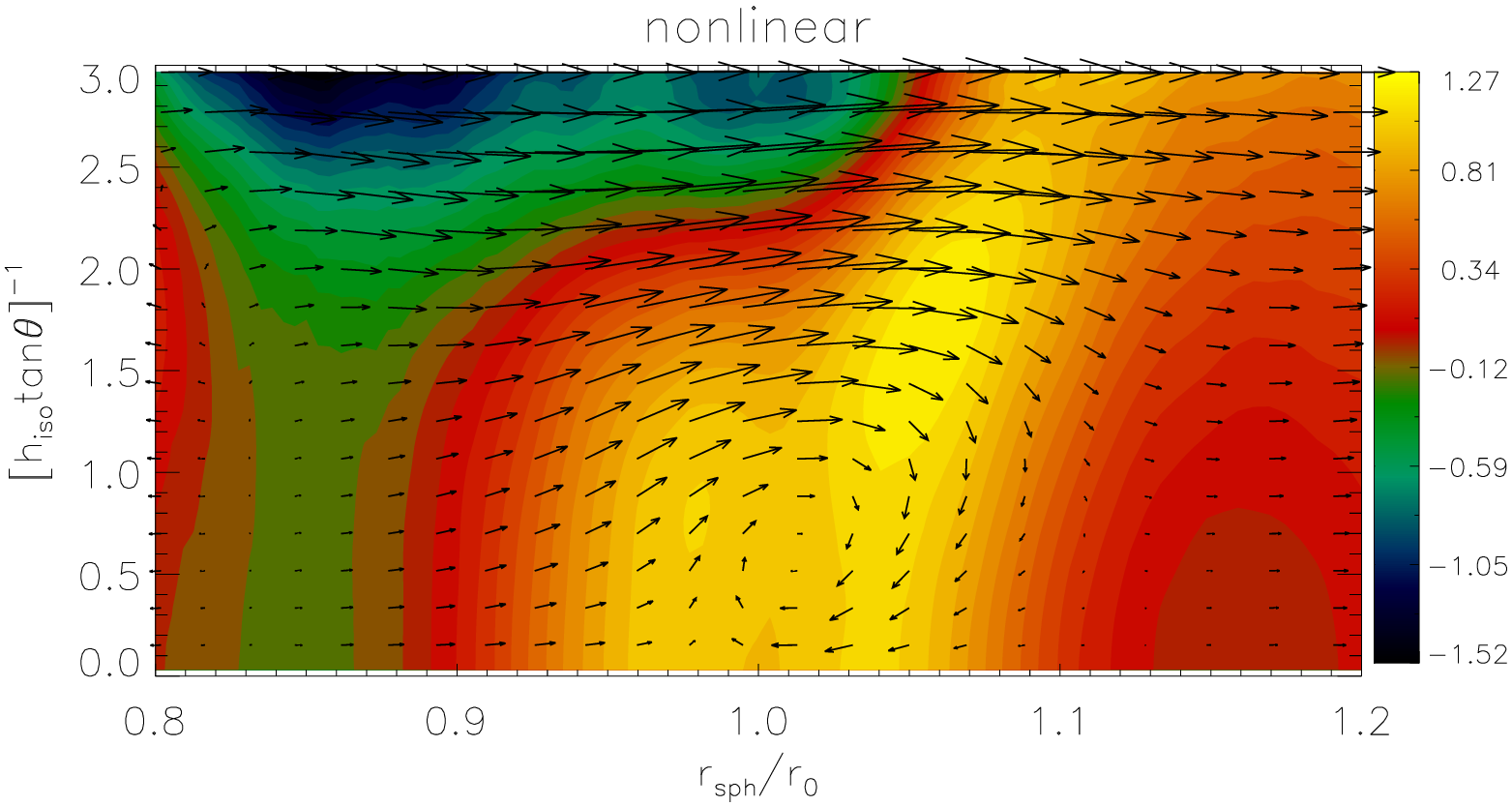}
  \includegraphics[scale=.425,clip=true,trim=0.0cm 
    .48cm 0.cm 0.4cm]{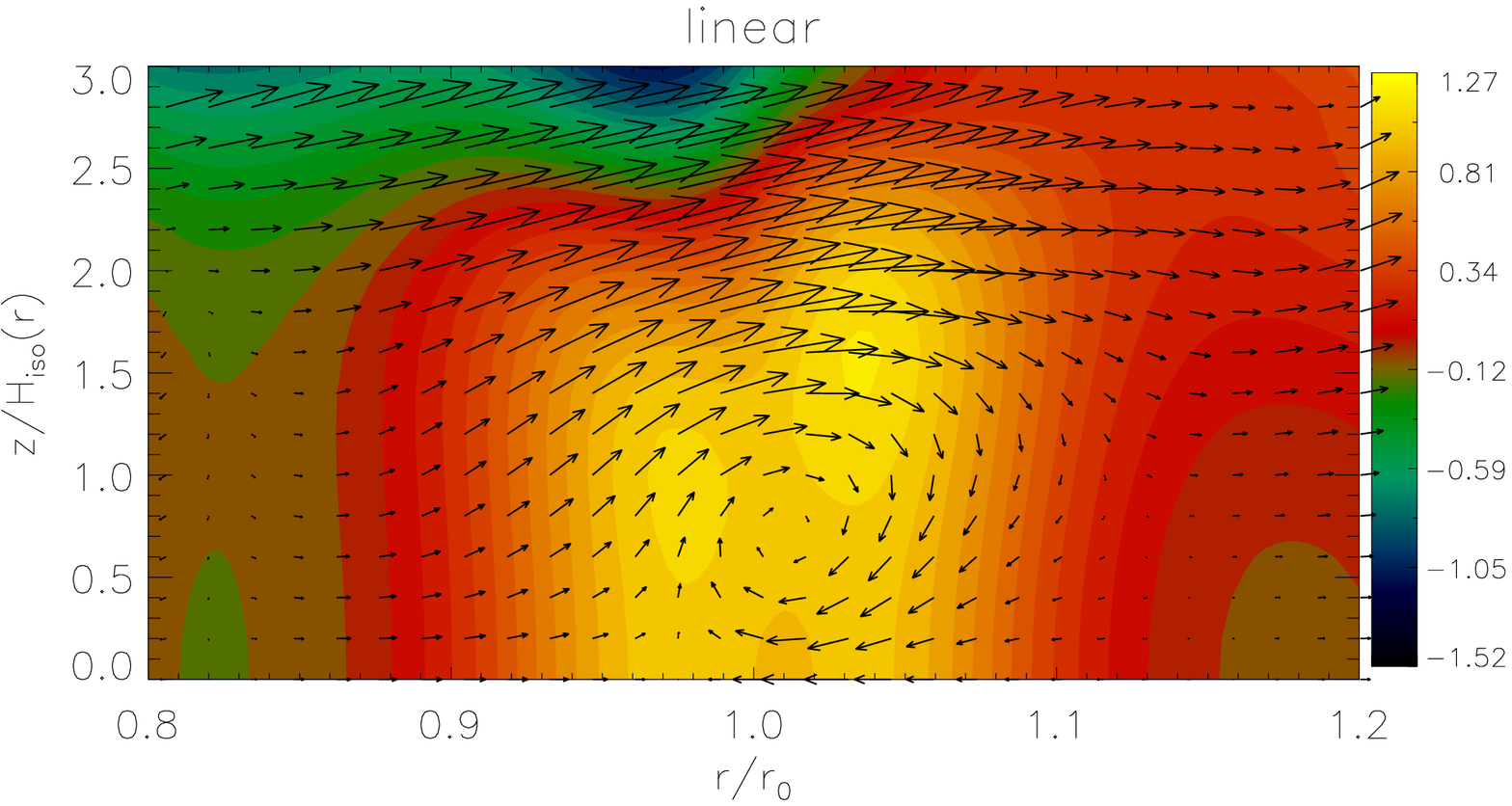}
  \caption{The perturbed velocity field projected onto the meridional
    plane at the vortex azimuth $\phi_0$, associated with the RWI
    calculated from a nonlinear hydrodynamic simulation (top) and the
    linear code (bottom). The average three-dimensionality, as
      measured by the ratio of vertical to meridional flow speeds,
      $\avg{\theta_m}$, is $0.39$ and $0.34$ in the linear and 
      nonlinear calculation, respectively. 
    A map of the normalized pressure
    perturbation is also shown. 
  \label{lin_nonlin_rz}}
\end{figure}

We compare meridional flows in Fig. \ref{lin_nonlin_rz}. The perturbed flow
is mostly horizontal in both cases.   
The nonlinear simulation also produce vortical motion in the same sense as
the linear calculation. For the \texttt{ZEUS} calculation, 
  we find the maximum magnitude of 
  vertical Mach number is $\sim 1\%$ with a density-weighted average
  value of $0.15\%$ in the shell $r_\mathrm{sph}\in[0.9,1.1]r_0$. 
The asymmetry of the pressure perturbation
about $r_0$ is captured by the linear code as well. Disagreement toward
the upper boundary is not unexpected, since the linear code assumes
the upper boundary is at a constant number of scale-heights above the
midplane, whereas the spherical grid imposes constant opening angle. 
However, both plots indicate $W$ increases away from the midplane in
the region exterior to $r_0$.

\section{Summary and discussion}\label{summary}
In this paper, we have examined the linear stability of radially
structured three-dimensional disks with non-uniform entropy
distribution. These calculations may be considered as an extension to 
the 2D Rossby wave instability \citep{li00} by adding the vertical
dimension, or to the barotropic RWI calculations of \citetalias{lin12} by
adding an energy equation with a simpler numerical method. 

We adopted polytropic disk equilibria so that
the magnitude of entropy gradients can be conveniently parametrized
by $\Delta\gamma\equiv\gamma/\Gamma$, and we focused on the effect of
$\Delta\gamma\geq1$. When the background density and
velocity field is fixed through $\Gamma$, we found increasing $\Delta
\gamma$ has negligible effect on the instability growth rate. However,
 the magnitude of pressure and density perturbations increase with height, and the 
meridional flow associated with the vortex core is qualitatively
changed, with the introduction of meridional vortical
motion. 

 Meridional vortical motion was found to correlate with a small 
  tilt of a fluid column with negative vertical vorticity
  perturbation.  In standard hydrodynamics, vorticity tilting
  can originate from a contribution of the
  $\bm{\omega}\cdot\nabla\bm{v}$ term in the evolution equation of the
  vorticity independently of the baroclinic source term $\nabla \rho
  \times\nabla p$. However, given the tilt is absent in our homentropic
  calculations, we associate the tilt with the baroclinic source term, which
  produces azimuthal vorticity. 
We also found
that the vertical velocity at the vortex core is no longer linear in
$z$, as for homentropic flow. 

In our second set of experiments, we fixed $\gamma$ and decreased
$\Gamma$. We found that by making the flow nonhomentropic, the
co-rotation region became \emph{more} three-dimensional, despite the
decrease in growth rate. This result
is opposite to \citetalias{lin12} where lowering $\Gamma$ made the
flow less three-dimensional. This implies that entropy gradients  
play an important role in the vertical structure of the perturbations. 
%This will likely have impact
%on the motion of dust particles within Rossby vortices, in particular 
%as the flow is dominated by bouyancy forces away from the midplane. 

We also considered isothermal equilibria. A linear calculation with 
$\Gamma=1.1$ and one with a strictly isothermal setup ($\Gamma\equiv
1$) were consistent. Both produced prominent meridional 
vortical motion. In order to verify this feature, we ran a nonlinear
simulation of the RWI in an initially isothermal disk, but evolved
adiabatically. We indeed identified said  vortical motion.    
%although the agreement with linear calculations is
%qualitative. This might imply that nonlinear effects become important
%for the RWI vertical structure earlier than in the horizontal
%direction. 
Keeping in mind that the setup for linear and nonlinear simulations 
were not identical (e.g. numerical grid, boundary treatment), similarities
between them, such as mode frequency, growth rate and horizontal flow,
are satisfactory.  

Vortical motion in the meridional plane thus appears characteristic of 
the \emph{linear} RWI in nonhomentropic disks. Whether or not this is 
significant for the vortex evolution can only be answered by detailed 
long term nonlinear simulations. If this vortical motion is present 
in the nonlinear regime then it may prevent dust particles from reaching the disk surface, 
which occurs for homentropic flow \citep{meheut12c}. 

However, given this meridional vortical motion is absent in the homentropic 
linear solution, it may eventually vanish because of entropy
mixing, if no mechanism is present to maintain entropy gradients. 
For example, the background entropy increases with height but 
the linear entropy perturbation becomes more negative with height,  
and its magnitude grows exponentially in time. 
 Indeed,  recent 3D fully compressible simulations in
  nonhomentropic disks shows that well into the nonlinear regime,
  Rossby vortices have columnar structure 
  \citep{richard13}. On the other hand, \cite{meheut12} observed
strong meridional vortical motion in their homentropic hydrodynamic
simulations; we conclude they are of nonlinear origin.  

In the linear solutions, we often observe perturbation magnitudes
increase away from the midplane in nonhomentropic
disks\footnote{This reminds us of the off-midplane vortices
  discovered by \cite{barranco05} in nonlinear local simulations, 
%  which suggest the non-zero vertical bouyancy frequency to be
%  stabilizing   against secondary instabilities \citep{lesur09}. 
  but the  
  setup considered in that study is very different from the
  present work. Nevertheless, in both cases the vertical entropy
  gradient is stabilizing away from the midplane. 
  %%  Nevertheless, a positive vertical bouyancy frequency
  %% may stabilize 3D vortices \citep{lesur09}.
} 
(e.g. Fig. \ref{lin_nonlin_rz}). Then the RWI may not be as robust
against vertical boundary conditions as it is to radial boundary
conditions. This could pose difficulty for the RWI to develop in dead
zones of real protoplanetary disks, which are expected to be confined
from above and below by magnetically turbulent layers
\citep{oishi09}. The vertical boundary condition set by these layers
may or may not be compatible with the linear RWI solution.

\subsection{Caveats and outlooks}
%need a good guess
One trade-off for the simplicity of our numerical method for 
linear simulations is that a trial eigenfrequency must be
guessed. This is not a significant obstacle for the problem at
hand, because previous RWI studies provide an important guide
\citep[][]{li00}. Otherwise, zeros of the complex function
$\mathcal{D}(\sigma)=\mathrm{det}\,\bm{U}$ need to be located with more 
rigorous methods \citep[e.g.][]{kojima86,valborro07}. We have also exploited
previous findings that the PPI and RWI are predominantly two-dimensional 
\citep{papaloizou85, goldreich86, kojima89, umurhan10, meheut12, lin12, lin12c}, which
enabled the use of a small number of basis functions. 
However, there could exist parameter regimes where the RWI
has significant vertical structure, rendering our solution method
inefficient. 

%extension to baroclinic equilibria 
%simple polytrope background
Our conclusions are limited to polytropic backgrounds. While this was 
convenient for numerical experiments, it is an over-simplification
of protoplanetary disks, which are 
expected to have complicated vertical structure \citep{terquem08}. In
particular, we found that entropy gradients plays a role in the
vertical structure of the linear RWI, and even a small entropy gradient can noticeably
modify the vertical flow (\S\ref{nz_nonzero}).
Thus, a realistic model for entropy evolution is needed. 
%This is important because even a small entropy gradient can noticeably 
%modify the vertical flow (\S\ref{nz_nonzero}). 

It would also be of interest to generalize the calculations
to \emph{baroclinic} equilibria\footnote{In fact, baroclinic tori were
  briefly considered by \cite{frank88}.}, 
for which $\p_z\Omega\neq0$. This may well be the 
case when the equilibrium pressure depends on both the density and temperature. 
%This requires 
%derivation of the linearized equations including terms involving $\p_z\Omega$. 
Complications from baroclinic instabilities may arise, however \citep{knobloch86, umurhan12, nelson12}.
%% This may involve adding heating or cooling
%% sources to the energy equation. 
%source terms - friction, cooling
%new physics - self-gravity

We have neglected gas self-gravity in this study. Our models therefore
assume that the Toomre parameter is much larger than unity in both the 
unperturbed \emph{and} perturbed states.  However, self-gravity
  may affect the RWI even when the Toomre parameter is not small
  \citep{lovelace12}. 
Previous studies
have found  higher $m$ RWI modes are favored when disk self-gravity is
included \citep{lyra08,lin11a}. Recent 3D simulations of the RWI in a
locally isothermal disk show that vertical self-gravity can noticeably
enhance the density perturbation near the midplane,
even though the initial disk was considered low mass \citep{lin12b}. %%  The effects
%%  of self-gravity %% should be considered in conjunction with disk
%%  thermodynamics. %In a nonlinear simulation, the latter may not hold even if the former
%does.
%Previous studies of the RWI also find higher $m$ modes modes are
%favored when disk self-gravity is included
%\citep{lyra08,lin11a}.    

In principle, one can express the Poisson integral as a matrix
operator and  incorporate it into our formalism. The linear problem is
further complicated by the need of a numerical solution to the
equilibrium equations describing a radially structured,
self-gravitating 3D disk \citep{muto11}.  Such a linear calculation is
beyond the scope of this paper, but will be inevitable for
understanding  the RWI in 3D self-gravitating disks. Perhaps a simpler
starting point, to gain first insight, is  direct hydrodynamic 
simulations including disk gravity. This is indeed the approach taken in our follow-up paper.    
%\clearpage
%since the ZEUS-MP code already includes a Poisson solver (as used in Lin 2012b). 

%have shown that
%self-gravity reduces the RWI growth rate through the potential perturbation, 
%but may be destabilizing through modification of the background disk \citep{lin11a}. 

\acknowledgments

I thank the referee, P. Barge, for suggesting the idea of a tilted vorticity column. 
I also thank S.-J. Paardekooper for comments on the first version of this paper. 

%\clearpage
\appendix
\section{PDE coefficients}\label{expressions}
In $(R,Z)$ co-ordinates, the coefficients for the PDE pair (Eq. \ref{contin_coeff_bar}---\ref{energy_coeff_bar}) 
with dependent variables $(\wbar,\qbar)$ are :
 
\begin{align}
 &a_1 = 1, \quad b_1 = -2Z\dbigH, \quad c_1 = Z^2\left(\dbigH\right)^2 - \frac{D}{\sbar^2H^2}, \quad
 d_1 = \left[\ln{\left(\frac{R}{D}\right)}\right]^\prime, \notag\\ 
 &e_1 = Z\left\{\left(\dbigH\right)^2 - \left(\dbigH\right)\left[\ln{\left(\frac{R}{D}\right)}\right]^\prime-
                 \left(\dbigH\right)^\prime\right\}, \quad
 f_1=\frac{2m\Omega}{R\sbar}\left[\ln{\left(\frac{\Omega}{D}\right)}\right]^\prime - \frac{m^2}{R^2},\notag\\
 & \bar{d}_1 = -\frac{1}{L_p}, \quad \bar{e}_1 = \frac{Z}{L_p}\dbigH + \frac{D}{\sbar^2HH_p},\notag\\
 &\bar{f}_1 = \frac{2m\Omega}{L_pR\sbar} - \frac{D}{c_s^2} + \frac{D}{\sbar^2 H}\frac{\p H_p^{-1}}{\p Z}
 - \frac{1}{L_p}\left[\ln{\left(\frac{R}{D}\right)}\right]^\prime - \frac{\p L_p^{-1}}{\p R} + Z\dbigH\frac{\p L_p^{-1}}{\p Z},
\end{align}
and
\begin{align}
&d_2 = \frac{\sbar}{L_sD},\quad e_2 = -\left(\frac{Z\sbar}{L_sD}\dbigH + \frac{1}{\sbar HH_s}\right), \quad
f_2 = \frac{2m\Omega}{L_sRD} - \frac{\sbar}{c_s^2} ,\notag\\
&\bar{f}_2 = \sbar\left( \frac{1}{c_s^2} - \frac{1}{DL_sL_p}\right) + \frac{1}{\sbar H_pH_s}. 
\end{align}
Note that these coefficients are expressed in terms of pressure,
entropy length-scales and the adiabatic sound speed. 
Although $H$ has
the physical meaning of the polytropic disk thickness, as far as the
derivation of these coefficients is concerned, it is simply a function
involved in a co-ordinate transformation. These expressions are
therefore valid for any barotropic equilibria.  

\section{Numerical route to a matrix equation for $W$}\label{alternative}
In \S\ref{linear} we arrived at the differential equation $UW=0$ by first deriving  
an equation for $\wbar$ then changed the dependent variable to $W$. 
Instead, we can first make the substitution $\wbar=\rho W$ and $\qbar=\rho Q$ 
in Eq. \ref{contin_bar}---\ref{energy_bar}, to obtain the 
governing equations for $(W,Q)$: 	
\begin{align}
  &A_1 \frac{\p^2 W}{\p R^2} + B_1\frac{\p^2W}{\p Z\p R} +
  C_1\frac{\p^2W}{\p Z^2} + D_1\frac{\p W}{\p R} + E_1\frac{\p W}{\p
    Z} + F_1W\notag\\
  & +\bar{D}_1\frac{\p Q}{\p R} + \bar{E}_1\frac{\p Q}{\p Z} +
  \bar{F}_1Q=0,\label{numerical_eq1}\\
  &D_2\frac{\p W}{\p R} + E_2\frac{\p W}{\p Z} + F_2 W + \bar{F}_2Q = 0.\label{numerical_eq2}
\end{align}
with
\begin{align}
&A_1 = a_1, \quad B_1 = b_1, \quad C_1 = c_1, \quad D_1 = 2a_1\Rp + b_1\Fp + d_1, \quad
E_1 = b_1\Rp + 2c_1\Fp + e_1,\notag\\
& F_1 = a_1\Rpp + b_1\Rp\Fp + c_1\Fpp + d_1\Rp + e_1\Fp + f_1,\notag\\
&\bar{D}_1 = \bar{d}_1,\quad \bar{E}_1 = \bar{e}_1 ,\quad \bar{F}_1 = \bar{d}_1\Rp + \bar{e}_1\Fp + \bar{f}_1,\notag\\
& D_2 = d_2,\quad E_2 = e_2, \quad F_2 = d_2\Rp + e_2\Fp + f_2,\notag\\
& \bar{F}_2 = \bar{f}_2. 
\end{align}
We recall the unperturbed density is $\rho=\rho_0(R)g(Z)$ and primes denote differentiation
with respect to the argument. These transformation formulae make no reference to a polytropic
background, so they are valid for any equilibrium density field separable in the above form, such as
an exponential atmosphere. 
%In operator form, we write
%\begin{align}
%  &U_1\wbar + \bar{U}_1\qbar = 0,\label{op2_eq1}\\
%  &U_2\wbar + \bar{U}_2\qbar = 0.\label{op2_eq2}
%\end{align}

When discretized, these equations have the matrix representation
\begin{align}
  &\bm{U}_1\bm{w} + \bar{\bm{U}}_1\bm{q} = 0,\label{op2_eq1}\\
  &\bm{U}_2\bm{w} + \bar{\bm{U}}_2\bm{q} = 0,\label{op2_eq2}
\end{align}
where $\bm{q}$ is the vector of pseudo-spectral coefficients for $Q$,  
i.e. $Q_i(Z)\equiv Q(R_i,Z)=\sum_{k=1}^{N_Z} q_{ki}\psi_k(Z/Z_s)$. The matrix representation
of $UW=0$ is then
\begin{align}
\left[\bm{U}_1 - \bar{\bm{U}}_1\left(\bar{\bm{U}}_2^{-1}\bm{U}_2\right)\right]\bm{w}\equiv\bm{U}\bm{w}=  \bm{0}. 
\end{align}
Note that we can divide Eq. \ref{numerical_eq2} by $\bar{F}_2$ before converting the operators 
to matrices. Then $\bar{\bm{U}_2}$ is a block diagonal matrix consisting only of 
the Chebyshev polynomials evaluated at vertical grid points. 
Its inverse can be pre-computed and stored.  

In this approach, the user only needs to specify the PDE coefficients 
defined in Appendix \ref{expressions}. % The above coefficient transformation can be coded in a subroutine. 
The transformed coefficients $A_1$---$F_1$ are used to construct the
matrix $\bm{U}_1$ as described in \cite{lin12c}, and similarly for
$\bar{\bm{U}}_1$ and $\bm{U}_2$. The final operator, $\bm{U}$, results from
matrix multiplication and addition, for which standard software can perform.  

\section{Estimating instantaneous mode growth rates}\label{instant_growth}
When dealing with hydrodynamic simulations it may be impractical to frequently output data for 
explicit computation of time derivatives. This is particular the case if high spatial resolution 
simulations are performed. However, we can take advantage of this and 
exchange time derivatives for spatial derivatives using the fluid equations. 

As usual, denote the Fourier transform with subscript $m$, so that
\begin{align}
 \rho_m(r,\theta, t) \equiv \int_0^{2\pi} \rho(r,\theta,\phi,t)\exp{(-\imgi m\phi)}d\phi,
\end{align}
where we have adopted spherical co-ordinates, so here $r$ is the spherical radius. Taking a time derivative and using
the continuity equation gives 
\begin{align}
 \frac{\p\rho_m}{\p t} = - \int_{0}^{2\pi} \nabla\cdot\left(\rho\bm{v}\right)\exp{(-\imgi m\phi)}d\phi. 
\end{align}
Writing this out in full, applying the usual rule for 
Fourier transforms to the azimuthal contribution to the divergence, we obtain
\begin{align}
 -\frac{\p\rho_m}{\p t} =  \frac{1}{r^2}\frac{\p}{\p r}\left[r^2\left(\rho v_r\right)_m\right]
                          + \frac{1}{r\sin{\theta}}\frac{\p}{\p \theta}\left[\sin{\theta}\left(\rho v_\theta\right)_m\right]
                          +\frac{\imgi m}{r\sin{\theta}}\left(\rho v_\phi\right)_m.
\end{align}
We can therefore just use the Fourier transform of momentum densities to calculate time derivatives of a Fourier mode.  
The complex frequency $\sigma$ is defined through $\p_t\rho_m = \imgi\sigma\rho_m$, from which we extract
the mode frequency $\omega$ and growth rate $\nu$. These are spatially-dependent when obtained from simulation data using 
the above procedure. So we average $\omega$  and $\nu$ over the $\theta$ domain and around co-rotation $r\in[0.8,1.2]r_0$. 
This gives an estimate of the instantaneous growth rate and pattern speed of a mode with azimuthal wavenumber $m$ at time $t$.

%\bibliographystyle{apj}
%\bibliography{ref}

\end{document}